\documentclass[journal]{new-aiaa}
\usepackage[utf8]{inputenc}
\usepackage{textcomp}

\usepackage{graphicx}
\usepackage{amsmath}
\usepackage[version=4]{mhchem}
\usepackage{siunitx}

\usepackage{bm}
\usepackage{amsfonts}

\usepackage{caption}
\usepackage{subcaption}
\usepackage{multicol}

\usepackage[verbose]{placeins}
\usepackage{soul}
\usepackage{adjustbox}
\usepackage{multirow}
\usepackage{array}
\newcolumntype{L}[1]{>{\raggedright\arraybackslash}p{#1}}
\newcolumntype{C}[1]{>{\centering\arraybackslash}p{#1}}
\usepackage{ulem}
\usepackage{nicefrac, xfrac}

\usepackage{graphicx,import}
\usepackage{pgfplots}
\usepackage{tikz}
\usetikzlibrary{positioning,external}
\pgfplotsset{compat=newest,
every plot/.append style={color=black, mark=none},
every axis/.append style={
label style={font=\fontsize{9pt}{1em}\color{white!15!black}\selectfont},
tick style={font=\fontsize{9pt}{1em}\selectfont, color=black, line cap=round},
ticklabel style= {font=\fontsize{9pt}{1em}\selectfont},
legend style={legend cell align=left, font=\fontsize{9pt}{1em}\selectfont, align=left, draw=white!15!black},
/pgf/number format/1000 sep={},
yticklabel style={
        /pgf/number format/fixed,
        /pgf/number format/precision=5
},
scaled y ticks=false
}}

\usepackage[capitalise]{cleveref}
\usepackage{algorithm}
\usepackage{algpseudocode}

\usepackage[nopostdot,acronym,nogroupskip,nonumberlist]{glossaries}
\newacronym{cr3bp}{CR3BP}{Circular Restricted Three-Body Problem}
\newacronym{leo}{LEO}{Low Earth Orbit}
\newacronym{bc}{BC}{Ballistic Capture}
\newacronym{etd}{ETD}{Energy Transition Domain}
\newacronym{zvc}{ZVC}{Zero Velocity Curve}
\newacronym{po}{PO}{Periodic Orbit}
\newacronym{dro}{DRO}{Distant Retrograde Orbit}
\newacronym{p3dro}{P3DRO}{Period-Tripled Distant Retrograde Orbit}
\newacronym{p4dro}{P4DRO}{Period-Quadruped Distant Retrograde Orbit}
\newacronym{qpo}{QPO}{Quasi-Periodic Orbit}
\newacronym{tof}{ToF}{Time of Flight}
\newacronym{da}{DA}{Differential Algebra}
\newacronym{ads}{ADS}{Automatic Domain Splitting}
\newacronym{nrho}{NRHO}{Near-Rectilinear Halo Orbit}
\newacronym{tpbvp}{TPBVP}{Two-Point Boundary Value Problem}
\newacronym{ic}{IC}{Initial Condition}
\newacronym{ode}{ODE}{Ordinary Differential Equation}
\newacronym{scp}{SCP}{Sequential Convex Programming}
\newacronym{ad}{AD}{Automatic Differentiation}
\newacronym{stm}{STM}{State Transition Matrix}
\glsdisablehyper

\usepackage{optidef}

\usepackage{enumitem}

\usepackage{lipsum}
\newcommand\blfootnote[1]{%
  \begingroup
  \renewcommand\thefootnote{}\footnote{#1}%
  \addtocounter{footnote}{-1}%
  \endgroup
}

\usepackage{amsmath}
\usepackage[version=4]{mhchem}
\usepackage{siunitx}
\usepackage{longtable,tabularx}
\newlength{\minuslength}
\usepackage{float}	 											
\usepackage[position=bottom,
			font=footnotesize,
		 	skip=0pt]{subcaption}								
\usepackage{setspace} 											
\usepackage{caption}											
\newcommand\newl[2]{ 											
 	\expandafter\newlength\csname #1\endcsname
 	\expandafter\setlength\csname #1\endcsname{#2}}
\newcommand\setl[2]{ 											
 	\expandafter\setlength\csname #1\endcsname{#2}}

\newl{figh}{.0\columnwidth}										
\newl{figw}{.0\columnwidth}										
\newl{boxh}{.0\columnwidth}										
\newl{boxw}{.0\columnwidth}										
\newl{tcw}{.0\columnwidth}										

\newl{subfigskip}{.6em}											
\newl{plotvsep}{.8em}											

\newcommand{\lastdate}{Jul 1, 2025}

\usepackage{pgfplots}
\usepgfplotslibrary{groupplots}
\usetikzlibrary{positioning}

\pgfplotsset{
	compat=newest,
	plot coordinates/math parser=false,
}

\pgfdeclarelayer{annotation}
\pgfdeclarelayer{grid}
	\pgfsetlayers{grid,main,annotation}

\title{Optimization of Transfers linking Ballistic Captures to Earth-Moon Periodic Orbit Families} 

\author{Lorenzo Anoè \footnote{PhD student, Te P\=unaha \=Atea -  Space Institute, University of Auckland, 20 Symonds Street, Auckland 1010, New Zealand. \\ Corresponding author. Email: \textit{lorenzo.anoe@gmail.com}} }

\author{Roberto Armellin \footnote{Professor, Te P\=unaha \=Atea -  Space Institute, University of Auckland, 20 Symonds Street, Auckland 1010, New Zealand.}}

\author{Jack Yarndley \footnote{PhD student, Te P\=unaha \=Atea -  Space Institute, University of Auckland, 20 Symonds Street, Auckland 1010, New Zealand.}}
\affil{The University of Auckland, Auckland, 1010, NZ}

\author{Thomas Caleb \footnote{PhD student, ISAE-SUPAERO, 10 avenue Marc Pélegrin, Toulouse, 31055, France.}}

\author{Stéphanie Lizy-Destrez \footnote{Full professor, ISAE-SUPAERO, 10 avenue Marc Pélegrin, Toulouse, 31055, France.}}
\affil{ISAE-SUPAERO, Toulouse, 31055, France}

\begin{document}

\maketitle

\begin{abstract}
The design of transfers to periodic orbits in the Earth–Moon system has regained prominence with NASA’s Artemis and CNSA’s Chang’e programs. This work addresses the problem of linking ballistic capture trajectories — exploiting multi-body dynamics for temporary lunar orbit insertion — with bounded periodic motion described in the circular restricted three-body problem (CR3BP). A unified framework is developed for optimizing bi-impulsive transfers to families of periodic orbits via a high-order polynomial expansion of the CR3BP dynamics. That same expansion underlies a continuous 
parameterization of periodic orbit families, enabling rapid targeting and analytic sensitivity. Transfers to planar periodic orbit families — such as Lyapunov L1/L2 and distant retrograde orbits (DROs) — are addressed first, followed by extension to spatial families — such as butterfly and halo L1/L2 orbits — with an emphasis towards near-rectilinear halo orbits (NRHOs). Numerical results demonstrate low-$\Delta v$ solutions and validate the method’s adaptability for designing lunar missions. The optimized trajectories can inform an established low‐energy transfer database, enriching it with detailed cost profiles that reflect both transfer feasibility and underlying dynamical relationships to specific periodic orbit families. Finally, the proposed transfers provide reliable estimates for rapid refinement, making them readily adaptable for further optimization across mission-specific needs.
\blfootnote{Presented as paper 23-0288 at the 2023 AAS/AIAA Astrodynamics Specialist Conference, Big Sky, Montana, USA, August 13-17 2023.} 
\end{abstract}

\section{Introduction}

There is renewed interest in lunar missions, primarily driven by NASA's Artemis program~\cite{artemis2020,artemis2018EM1} and CNSA's Chang'e missions~\cite{Change_5T1}. Upcoming missions are expected to utilize a variety of operational orbits, some of which are naturally described within the \gls{cr3bp} framework. Notable examples include the CAPSTONE mission~\cite{Capstone2021}, which is currently testing the dynamics of a \gls{nrho}, and the \glspl{dro} employed by Artemis I~\cite{artemis2018EM1}.

However, the absence of closed-form solutions arising from the combined gravitational influence of the Earth and Moon — and further complicated by solar perturbations — poses significant challenges to the design of optimal cislunar missions. Addressing these challenges requires a detailed understanding, which is usually tackled through the use of the \gls{cr3bp}, a widely adopted model in astrodynamics.
As a Hamiltonian system, the \gls{cr3bp} conserves total energy, typically expressed via the Jacobi constant. This constant serves as a parameter for the generation of continuous families of \glspl{po}, including the planar Lyapunov families, the \gls{dro} family, and its period-tripling bifurcations in the \gls{p3dro} family. These families were first introduced (even though in the Hill problem) by Broucke and Hénon over 50 years ago~\cite{Broucke1968, henon1969V}. Since then, numerous additional families have been studied, such as the $L1$ and $L2$ halo orbits~\cite{howell1984}, and the butterfly family originating from the P2HO1 bifurcation~\cite{zimovan2020}.
A recent comprehensive study analyzes the global structure, bifurcations, and interconnections of many spatial \gls{po} families~\cite{aydin2025POfamilies}. The stability of these \glspl{po} is assessed using the monodromy matrix and its Floquet multipliers, while Poincaré section techniques reduce the planar dynamics to area-preserving maps~\cite{winter2000stabilityDRO, Capdevila-Guzzetti-Howell2014}, revealing local regions of stability known as \gls{dro} stability regions~\cite{Markellos1974, Douskos-Markellos2007}.

In this work, the target \glspl{po} are the \glspl{dro}, Lyapunov L1/L2, halo L1/L2 (which include \glspl{nrho}), and butterfly families.  
These families are operationally relevant because they offer complementary features for lunar mission design: \glspl{nrho} provide good stability together with close periapsis passages useful for staging and surface access~\cite{whitley2018-NRHOandButterflyForLunarExploration}, while higher-altitude arcs simplify Earth–Moon transfers~\cite{foust2019gateway, zimovan2020}. 
Butterfly orbits share some geometric properties of the orbits, but not their stability. By contrast, \glspl{dro} furnish broad low-energy transfer corridors and long-term stability, and Lyapunov/halo orbits act as natural transfer gateways and phasing options for cislunar operations.  
These complementary dynamical and operational properties explain the growing interest in these families. Beyond Earth–Moon applications, \glspl{po} have also informed mission design in other systems, including three-dimensional \glspl{po} around Phobos~\cite{marmo2021DROmmx}, sticky \glspl{dro} transfers in the Sun–Earth system~\cite{scott2010sticky}, and stability studies in the Jupiter–Ganymede system~\cite{Li-Tao-Jiang2022}.

This study focuses on the design of transfers to \glspl{po} in the Earth–Moon system, leveraging \gls{bc} trajectories as a staging point for future lunar missions. 
A \gls{bc} provides natural transport by exploiting the combined gravitational influence of two or more bodies to achieve a temporary, unpowered capture around the Moon. Recent work by Anoè \textit{et al.}~\cite{BC-ETDjournal, BC-3Djournal} developed practical methods to generate \glspl{bc} in both the planar and spatial \gls{cr3bp} for the Earth–Moon system. Because \glspl{bc} are inherently transient, small corrective maneuvers are required to transition a spacecraft into a bounded lunar orbit. Rather than aiming for direct insertion into low lunar orbits, this study examines bi-impulsive transfers that deliver a spacecraft from a given \gls{bc} into an operational \gls{po}. We show that, under the considered low-energy scenario, such transitions can be achieved with low $\Delta v$.  
Importantly, \glspl{bc} are reachable from \gls{leo} in low-energy mission architectures (for example, the \textit{Lunar Trailblazer} concept), where sequences of phasing maneuvers and lunar gravity assists place the spacecraft on a \gls{bc}-like approach to the Moon~\cite{BC-3Djournal}\footnote{\url{https://www.jpl.nasa.gov/news/how-nasas-lunar-trailblazer-will-make-a-looping-voyage-to-the-moon/}}. Such architectures exploit the weak stability boundary to reduce launcher/injection energy at the expense of longer flight times~\cite{Belbruno-Topputo-2, topputo_optimal2LunarTransfers4BP}. In this mission context, the BC phase serves as an integral component of the low-energy architecture, enhancing the flexibility of the mission design. It enables a wider range of lunar insertion options, lowers the associated maneuver cost, and offers alternative or contingency pathways for achieving capture. Accordingly, the $\Delta v$ values reported in this paper quantify the primary maneuver cost required for lunar insertion from the \gls{bc} (therefore excluding launch/injection, which is assumed to be provided by the launcher).

Target \glspl{po} are selected from a continuously parameterized family of \glspl{po} computed by Caleb et al.~\cite{caleb2023}, described through a high-order polynomial representation obtained via \gls{da}. This representation, referred to as an abacus, enables efficient access to \glspl{po} across the family through a compact, complete, and differentiable formulation. At the time of writing, the abacus provides access only to the Lyapunov $L1$/$L2$, \gls{dro}, halo $L1$/$L2$, and butterfly families. This constitutes the main limitation of the current work, which could be further strengthened by extending the target set to include additional three-dimensional \glspl{dro}~\cite{lara_classificationDSR_DROspatialFam} and other \gls{po} families~\cite{henon1969V, aydin2025POfamilies}.

Crucially, the same \gls{da} framework used to generate the \gls{po} abacus is also employed in the transfer optimization process. By leveraging high-order expansions of the dynamics, we develop a unified method that consistently exploits the benefits of \gls{da}-based techniques, such as rapid evaluation, local accuracy, and efficient sensitivity analysis. The resulting formulation enables the optimization of bi-impulsive transfers across the entire \gls{po} family, minimizing the total maneuver cost $\Delta v$.
Optimization variables include the initial phase along the \gls{bc}, the arrival phase on the \gls{po}, the parameter identifying a specific member of the \gls{po} family, and the \gls{tof}.

Using this optimization setup, transfers from \glspl{bc} to \glspl{po} are computed across various scenarios in both the planar and spatial \gls{cr3bp}. The method’s flexibility is demonstrated by the consistency and diversity of viable solutions, enabling reliable generation of connections between the proposed \gls{bc} trajectories and the wide range of target \glspl{po} families.
Notably, this represents a major advancement, as no previous work has demonstrated the capability to compute transfers from \glspl{bc} to arbitrary \gls{po} families. The high-order formulation based on \gls{da} enables an analytical representation of both the dynamics and the full set of target families through patched polynomial maps, allowing for a systematic exploration of the entire variable domain and a quantitative assessment of the dynamical relationships between \glspl{bc} and nearby \glspl{po}.
This capability significantly enhances the utility of the existing low-energy trajectory database~\cite{BC-ETDjournal, BC-3Djournal}, providing mission designers with precise estimates of the transfer cost to specific \glspl{po} and supporting a deeper investigation into fundamental questions such as: To what extent can the existence of \glspl{bc} be attributed to the dynamics of \glspl{po} invariant manifolds? And if such a connection exists, which \gls{po} family influences each \gls{bc}, and at what stage along its trajectory?

Finally, the optimality of the bi-impulsive solutions is validated using convex optimization techniques, as initially demonstrated by Jacini \textit{et al.}~\cite{IppolitaIAC}, who also explored their application in preliminary refinement processes. Building on this foundation, the present work incorporates refinement procedures developed by Yarndley \textit{et al.}~\cite{Yarndley2025, HoltISSFD}, further confirming the suitability of the computed bi-impulsive transfers as high-quality initial guesses for free-time, multi-impulsive optimization. While related studies have shown that bi-impulsive solutions can serve as good initial guesses for three-impulse transfers~\cite{grossi_optimal3LunarTransfers4BP}, our analysis extends these scenarios by quantitatively assessing the suboptimality of bi-impulsive solutions and providing additional insights into their suitability for multi-impulsive and potentially low-thrust or higher-fidelity optimization problems. We focus here on the L1 Halo family as a representative case study, which is particularly relevant for mission applications such as the Lunar Gateway and Lunar Trailblazer~\cite{BC-3Djournal}. These results, therefore, extend previous findings and demonstrate the practical value of the proposed bi-impulsive solutions in current mission design contexts.


The paper is structured in an incremental manner. It first introduces the relevant dynamical systems concepts and employs a Poincaré section analysis in the planar problem to build a clear understanding of the underlying dynamics and problem setup. Insights from this analysis are then used to initialize and solve the planar transfer optimization, which is subsequently extended to the spatial case. This stepwise approach facilitates the reader’s comprehension of both the dynamical background and the proposed optimization framework. The paper is organized as follows. \Cref{sec: background} introduces the core concepts and tools.
The relationship between \glspl{bc} and \glspl{po} is preliminarily investigated in \cref{sec: relationship BCs and POs and mono-imp} through Poincaré sections.
\Cref{sec: opt section} presents the formulation of the planar bi-impulsive transfer optimization, structured into three enveloping algorithms to ensure comprehensive coverage of the optimization variable space.
Subsequently, \cref{sec: planar results sec} reports the results for in-plane transfers, including sample trajectories from multiple \glspl{bc} targeting \gls{dro} and Lyapunov families.
\Cref{sec: spatial problem} extends the methodology to the spatial case, adapting the planar formulation and presenting optimized transfers connecting spatial \glspl{bc} to halo and butterfly families.
Finally, the convex-based refinement procedure and its results are discussed in \cref{sec: refinement convex}, followed by conclusions in \cref{sec: conclusions}.

\section{Background} \label{sec: background}
In this section, the fundamental concepts and tools underpinning this study are presented. It begins with the equations of motion for the \gls{cr3bp}, which provide the theoretical foundation for the model. The high-order expansion technique based on \gls{da} is then introduced, followed by a description of the abacus of \glspl{po} from~\cite{caleb2023} and the procedure used to generate the \glspl{bc}~\cite{BC-ETDjournal, BC-3Djournal}.

\subsection{Circular Restricted Three-Body Problem} \label{sec: cr3bp}
The \gls{cr3bp} is a fundamental model in celestial mechanics 
that describes the motion of a spacecraft $M_3$ under the gravitational influence of two celestial bodies $M_1$ and $M_2$, called primaries and with mass $m_1$ and $m_2$, respectively. The mass of $M_3$ is assumed to be negligible ($m_3 \ll m_1, m_2$), and gravitational parameters can be defined as $\mu_1 = G m_1$ and $\mu_2 = G m_2$, where $G$ is the universal gravitational constant. The mass ratio is therefore defined as
\begin{equation}
    \mu = \dfrac{\mu_2}{\mu_1+\mu_2} \, .
    \label{eq: mass ratio}
\end{equation}
The synodic frame has its origin at the system barycenter and rotates with the $M_1$–$M_2$ line; in this frame, the primaries remain fixed at $(-\mu,0)$ and $(1-\mu,0)$.

We nondimensionalize using the Earth–Moon distance (LU) and the Moon’s period (TU); derived units follow (Table \ref{tab: scaling units}).
The spacecraft state in the synodic frame at time $\tau$ is $\mathbf{x}(\tau) = (x,y,z,\dot x,\dot y,\dot z)$.
The pseudo-potential function is 
\begin{equation}
    \Omega = \dfrac{1}{2} \left( x^2 + y^2 \right) + \dfrac{1-\mu}{r_1} + \dfrac{\mu}{r_2} \, ,
\end{equation}
and hence the equations of motion for the satellite are
\begin{equation}
    \begin{cases} 
    \ddot{x} - 2\dot{y} = \dfrac{\partial \Omega}{\partial x} = x - (1-\mu)\dfrac{x+\mu}{r_1^3} - \mu \dfrac{x-(1-\mu)}{r_2^3} \\
    \ddot{y} + 2\dot{x} = \dfrac{\partial \Omega}{\partial y} = y - (1-\mu)\dfrac{y}{r_1^3} - \mu \dfrac{y}{r_2^3} \\
    \ddot{z} = \dfrac{\partial \Omega}{\partial z} = - (1-\mu)\dfrac{z}{r_1^3} - \mu \dfrac{z}{r_2^3}
    \end{cases}
    \label{eq: equations of motion}
\end{equation}
where $r_1$ and $r_2$ denote the distances from $M_3$ to $M_1$ and $M_2$, respectively
\begin{multicols}{2}
\noindent
\begin{equation}
r_1 = \sqrt{(x+\mu)^2+y^2+z^2}
\end{equation} 
\noindent
\begin{equation}
r_2 = \sqrt{[x-(1-\mu)]^2+y^2+z^2} \, .
\label{eq: r_2}
\end{equation}
\end{multicols}

\begin{table}
\caption{\label{tab: scaling units} Approximate scaling units used in this work for the Earth-Moon system.}
\centering
\begin{tabular}{L{1.5cm} L{3.5cm} L{4.5cm} L{4.8cm}}
\hline
\noalign{\vskip\doublerulesep}
Unit & Symbol & Value & Note \\ \hline
\noalign{\vskip\doublerulesep}
\noalign{\vskip\doublerulesep}
- & $\mu$ & $0.012150584269940$ & Mass ratio (see \cref{eq: mass ratio}) \\
\noalign{\vskip\doublerulesep}
\noalign{\vskip\doublerulesep}
Mass & $MU=G(m_1+m_2)$ & $4.035032\cdot10^{5} \; km^3/ s^2$ & System gravitational constant \\
\noalign{\vskip\doublerulesep}
\noalign{\vskip\doublerulesep}
Length & $LU$ & $384399 \; km$ & Mean Earth-Moon distance \\
\noalign{\vskip\doublerulesep}
\noalign{\vskip\doublerulesep}
Time & $TU=\left( LU^3/MU \right)^{0.5}$ & $2.357381 \cdot 10^6 \; s \approx 27.3 \, \text{days}$ & Moon's mean revolution period \\
\noalign{\vskip\doublerulesep}
\noalign{\vskip\doublerulesep}
Velocity & $VU=2\pi LU/TU$ & $1.024548 \; km/s$ & Mean orbital velocity of the Moon \\
\noalign{\vskip\doublerulesep}
\noalign{\vskip\doublerulesep}
Energy & $EU=VU^2=MU/LU$ & $1.049699 \; km^2/s^2$ & 
Moon's Keplerian energy \\
\noalign{\vskip\doublerulesep}
\hline
\end{tabular}
\end{table}

Five equilibrium points and an integral of motion are defined in this Hamiltonian system. They are referred to as Lagrange points ($L1$, $L2$, $L3$, $L4$, $L5$), and Jacobi constant $C_J$, respectively. The latter is defined as the sum of the kinetic $\mathcal{K}$ and potential $\Omega$ terms, and it reads
\begin{equation}
    C_J = -2 \left(\mathcal{K} - \Omega \right) = -\left(\dot{x}^2 + \dot{y}^2 + \dot{z}^2\right) + \left( x^2 + y^2 \right) + 2 \left( \dfrac{1-\mu}{r_1} + \dfrac{\mu}{r_2} \right) \, .
\end{equation}
As a Hamiltonian system, the \gls{cr3bp} admits continuous families of \glspl{po}, each parametrized by its Jacobi constant $C_J$.
When $C_J < C_J^{L1}$, the \glspl{zvc}~\cite{BattinZVC} open at $L1$, enabling transport feasibility between the region around $M_1$ and $M_2$.
A three-body energy parameter is defined as~\cite{BC-ETDjournal}
\begin{equation}
    \Gamma=\frac{C_{J}-C_{J}^{L1}}{C_{J}^{L4}-C_{J}^{L1}} \, .
    \label{eq: gamma}
\end{equation}
so that $\Gamma = 0$ when the $L1$ opening, and $\Gamma=1$ when the forbidden regions disappear ($C_{J}=C_{J}^{L4}$).

Finally, the two-body (Keplerian) energy with respect to $M_2$ reads
\begin{equation}
\varepsilon_{2} = \cfrac{v_{2}^{\,2}}{2}-\cfrac{\mu}{r_{2}} = 0 \, ,
\label{eq: vis viva}
\end{equation}
where the velocity $v_2$ is measured in the inertial frame of the second primary $M_2$ (i.e. the Moon).

\subsection{Differential Algebra} \label{sec: da intro}
The \gls{da} framework is a mathematical method used for the automatic expansion of sufficiently differentiable functions as a polynomial by replacing usual floating point operations with corresponding \gls{da} operations on a computer.

More specifically, \gls{da} technique is based on replacing a function $f$ with the map $\mathcal{M}$, which is the Taylor expansion of $f$ at order $k$~\cite{berz1999}. This approach allows for efficient computations and yields a polynomial representation of the function $f$ in a domain that can be easily estimated~\cite{wittig2015HOTM}. Additionally, the DA framework ensures well-defined algebraic and functional operations, as well as the composition inverse~\cite{berz1992}.
A key advantage of this method lies in the computation of the polynomial map only once, which can subsequently be evaluated at an arbitrary number of points. In other words, for calculating $S$ points, a single map generation is sufficient, followed by $S$ polynomial evaluations. In contrast, point-wise methods necessitate $S$ separate computations, as highlighted by Armellin \textit{et al.}~\cite{armellin2010AsteroidCloseEncounters}. The DA engine employed in this study is the Differential Algebra Core Engine (DACE), developed by Politecnico di Milano~\cite{rasotto2016DAtoolbox, massari2018DAsoftware}.
More details regarding the application of \gls{da} are provided in Sections~\ref{sec: relationship BCs and POs and mono-imp} and \ref{sec: opt section}.

\subsection{Abacus of POs} \label{sec: abacus of POs}
High-order polynomials have proven effective in mapping \gls{po} families of the \gls{cr3bp}, as presented in Caleb \textit{et al.}\cite{caleb2023}.
This technique enabled the possibility to generate an abacus of \gls{po} families, where evaluating polynomial maps in a 2D space $\left(p, \varphi\right)$ allows to determine the state and period $(\mathbf{x},T)$ that satisfy periodicity with a specified tolerance $\epsilon$, such that $\|\mathbf{x}(T)-\mathbf{x}(0)\|<\epsilon$.
The two coordinates of the mapping serve distinct purposes: the first enables users to select an orbit within the family using a parameter $p$ 
, while the second $\varphi\in[0,2\pi]$ corresponds to the phase on the \gls{po}. The variable $\varphi=0 \lor \varphi=2\pi$ usually indicates a state with $x<1-\mu$, $y=0$, $z=0$, $\dot{x}=0$, $\dot{y}\geq0$, and/or $\dot{z}\geq0$, depending on the specific family addressed.
For example, when evaluating a map from the abacus at coordinates $\left(p, \varphi\right)$, the result provides the position, velocity, and period $T$ of a member of the family at the given parameter $p$, after a time of $\frac{\varphi}{2\pi}\, T$ has elapsed from the $\varphi=0$ condition, namely
\begin{equation}
    \left( x, y, z, \dot{x}, \dot{y}, \dot{z}, T \right) = \mathcal{M}_{PO}(p,\varphi) \, .
    \label{eq: abacus map}
\end{equation}

To cover the entire domain of an abacus, the parameter space is partitioned into $K$ subdomains using \gls{ads}~\cite{wittig2015propagation}. Each subdomain, indexed by $k$, is approximated by a distinct polynomial map $\mathcal{M}^k_{PO}$ centered at an expansion point $(p_{k,c}, \varphi_{k,c})$. Each map ensures a prescribed level of precision within the parameter intervals $I_{p_k} = [p_{k,l} \, , p_{k,u}]$ and $I_{\varphi_k} = [\varphi_{k,l} \, , \varphi_{k,u}]$, where the subscripts $l$ and $u$ denote the lower and upper bounds of the interval.
A key advantage of this formulation is that it allows algebraic evaluation of the polynomial representations — without further propagation — to recover both the state and its derivatives with respect to $p$ and $\varphi$. This is also valid for any \gls{da}-represented function, as previously introduced in \cref{sec: da intro}.

Six families were mapped in the Earth-Moon system: the halo family at $L1$ and $L2$~\cite{howell1984}, the so-called ``butterfly'' family that originates from the P2HO1 bifurcation of the $L2$ halos~\cite{zimovan2020}, the planar Lyapunov orbits at $L1$ and $L2$ respectively known as the $G$ and $I$ families in Broucke~\cite{Broucke1968}, and the \gls{dro}, also referred to as the f family in Hénon~\cite{henon1969V}. 
These files are publicly available\footnote{Publicly available on Zenodo at the identifier: \url{https://doi.org/10.5281/zenodo.6778146} [last accessed \lastdate{}].}
and can be read using the C++ library DAHALOa\_reader\footnote{Library available at: \url{https://github.com/ThomasClb/DAHALOa_reader} [last accessed \lastdate{}].}.

\subsection{Ballistic Capture set generation} \label{sec: BC introduction}
\gls{bc} is a phenomenon by which a spacecraft or celestial body initially distant and outside the influence of a primary body is naturally transferred, under certain circumstances, to a temporary orbit around it (here referred to as capture).
More specifically, the initial distance must be comparable with the primaries' distance~\cite{Belbruno-Topputo-2}, and the successive capture phase must consist of at least one full revolution around $M_2$ (in its inertial frame) with negative two-body energy $\varepsilon_{2} < 0$. 
The detailed definition of \gls{bc} employed in this work is described in~\cite{BC-ETDjournal}, and it is only briefly summarized here for the readers' convenience.

Recently, a method of generating \glspl{bc} was developed by Anoè et al.~\cite{BC-ETDjournal} in the planar \gls{cr3bp}. In a later work, the method was extended to the spatial \gls{cr3bp}, enabling the creation of a database of spatial \glspl{bc}~\cite{BC-3Djournal}. In these works, \glspl{bc} were identified and analyzed across different celestial systems using the concept of the \gls{etd}.
In particular, the main feature of the \gls{etd} is that it constrains the value of the Jacobi constant and imposes zero two-body energy, namely $\varepsilon_{2} = 0$. 
In the planar \gls{cr3bp}, these two constraints reduce the four degrees of freedom into two degrees of freedom that can be represented in the synodic frame, hence enabling an analytical computation of an initial velocity for every initial position. The domain where this initial velocity is defined is called \gls{etd},
and it can be used as a fundamental tool to target the region of the synodic plane from which \gls{bc} trajectories emanate.
This approach has been applied to the Earth–Moon system (and others), demonstrating the ability to efficiently generate an exhaustive capture set $\mathcal{C}(\Gamma)$ of initial conditions leading to \glspl{bc}, stored together with relevant trajectory data in a structured database~\cite{BC-ETDjournal}. The subsequent spatial extension~\cite{BC-3Djournal} generalized this framework by introducing two additional variables, the out-of-plane coordinate $z$ and its associated velocity $\zeta$, thereby defining a three-dimensional capture set $\mathcal{C}(\Gamma, z, \zeta)$. The spatial method follows the same principles but adapts the detection strategy to include out-of-plane dynamics, resulting in a comprehensive database of spatial \glspl{bc}. In the following, for clarity, we introduce the formulation in its planar form only, while referring to~\cite{BC-3Djournal} for further details and to \cref{sec: spatial problem} for its application to the spatial case.

\begin{figure}[tb]
    \begin{subfigure}[t!]{0.49\textwidth}
        \centering
        \includegraphics[width=\textwidth]{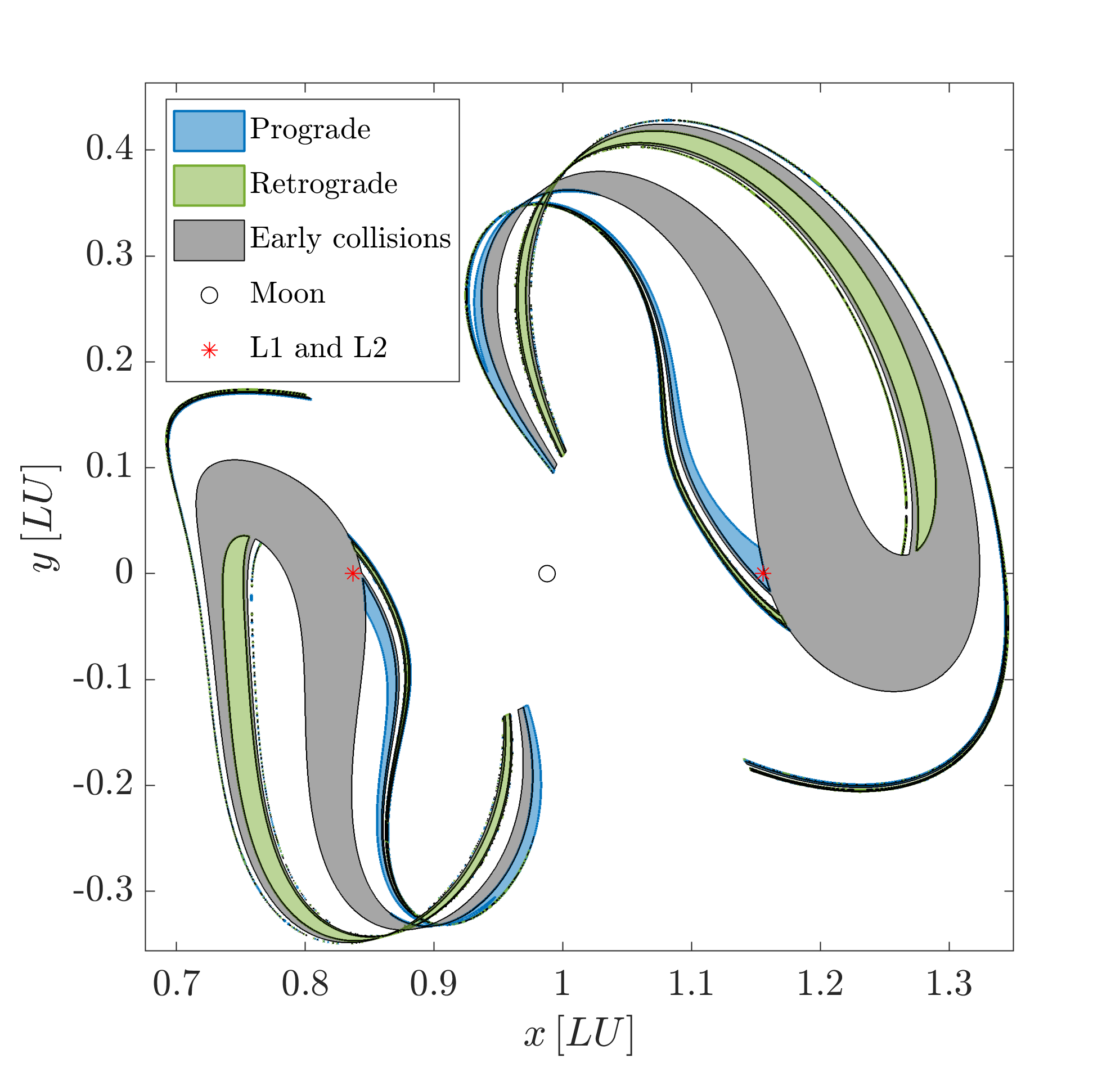}
        \subcaption{$\mathcal{C}(\Gamma = 0.84)$ early collisions, prograde and retrograde \glspl{bc}}
        \label{fig: capture set}
    \end{subfigure}
    \begin{subfigure}[t!]{0.49\textwidth}
        \includegraphics[width=\textwidth]{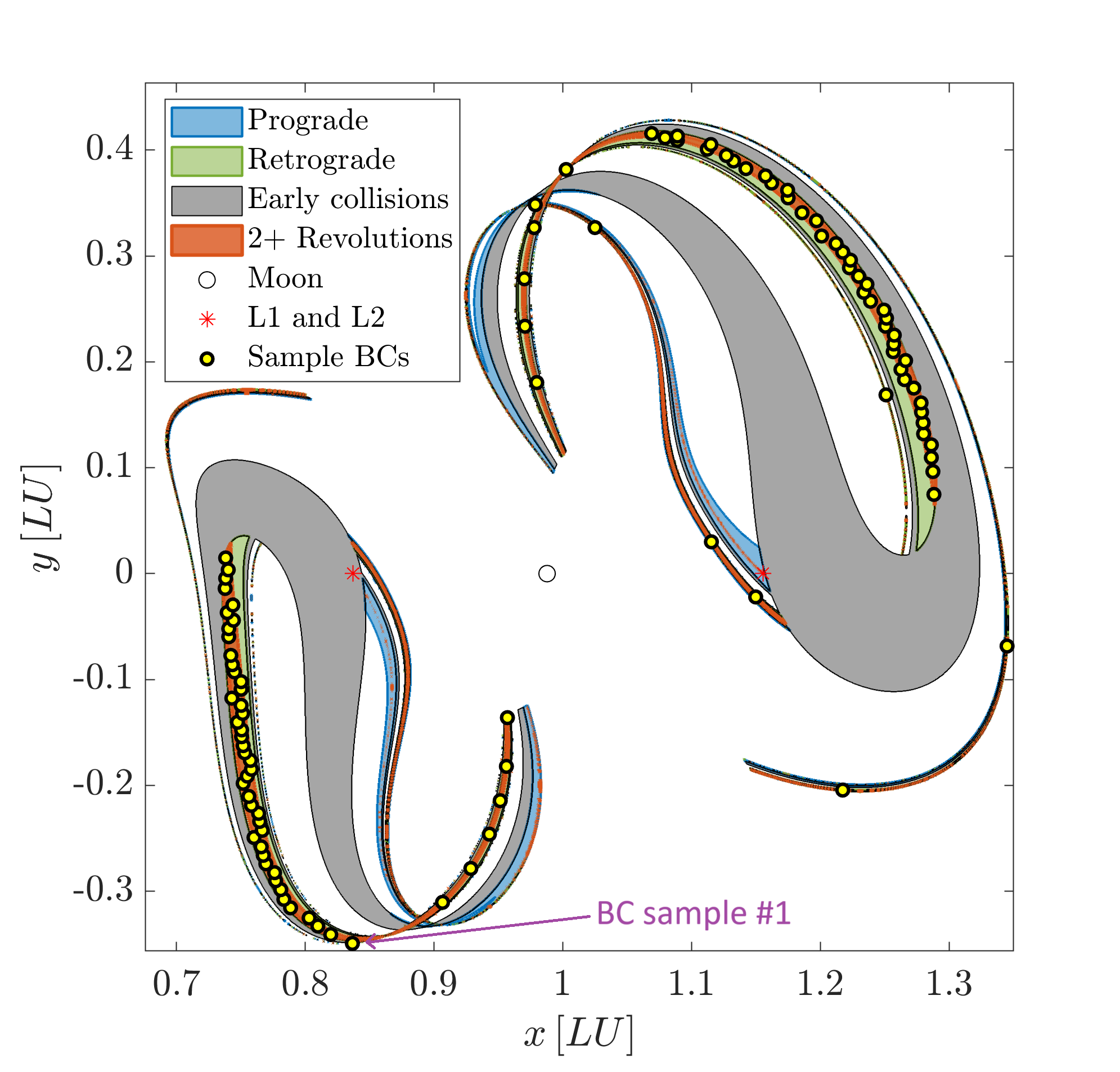}
        \subcaption{Sample \glspl{bc} representative of the 2+ retrograde capture subset} 
        \label{fig: sample BCs on capture set}
    \end{subfigure}
    \caption{Capture set $\mathcal{C}(\Gamma = 0.84)$ from~\cite{BC-ETDjournal}.} 
    \label{fig: capture set and sample BCs}
\end{figure}

An example of the planar capture set $\mathcal{C}(\Gamma = 0.84)$~\cite{BC-ETDjournal} for a fixed value of $\Gamma=0.84$ is represented in \cref{fig: capture set}. Note that the other two variables are assumed to be $z=0$ and $\zeta=0$ in the planar problem, therefore they are not indicated explicitly when describing $\mathcal{C}$. This includes initial conditions of trajectories leading to \gls{bc} when propagated (forward for the capture phase and backward for the escape leg). Prograde and retrograde \glspl{bc} are especially indicated. In \cref{fig: sample BCs on capture set}, only \glspl{bc} completing two or more revolutions are highlighted in orange. This allows for a selection of suitable trajectories for insertion into the \gls{dro} family. A representative sampling of the capture subset containing 2 or more retrograde revolutions is represented by yellow, black-contoured circles. A total of 104 \glspl{bc} are extracted to span the entire subset uniformly, representing $0.01\%$ of the \glspl{bc} in the aforementioned subset.

\begin{figure}[tb]
    \centering
    \begin{subfigure}[t!]{0.41\textwidth}
        \centering
        \includegraphics[width=0.97\textwidth]{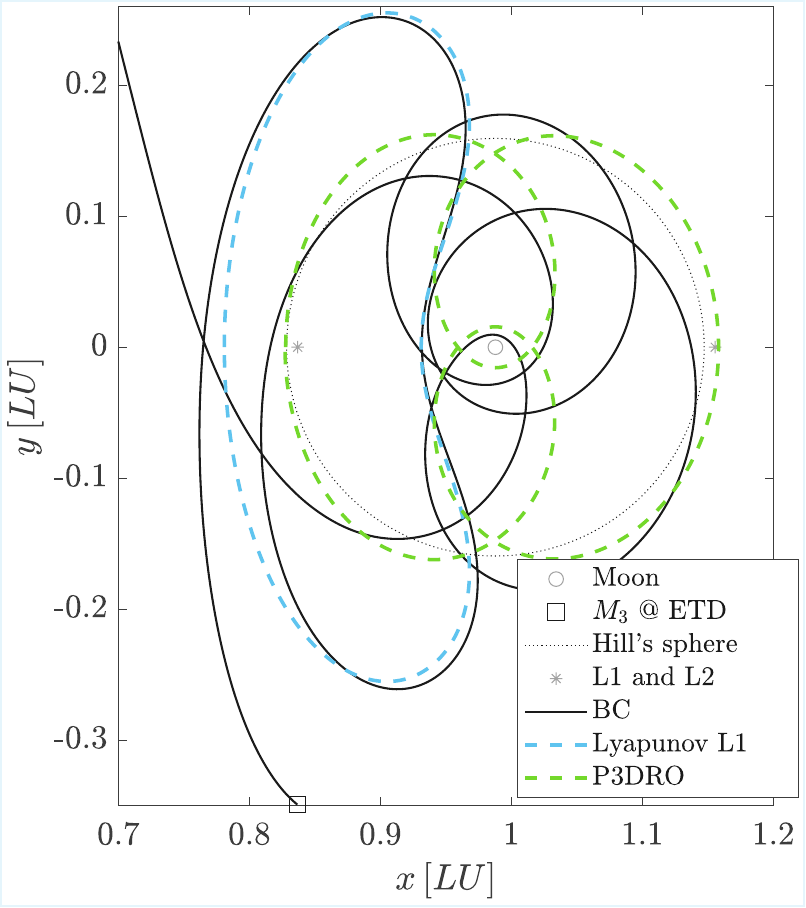}
        \subcaption{Sample \gls{bc}~$\#1$ trajectory from $\mathcal{C}(\Gamma = 0.84)$ of \cref{fig: capture set and sample BCs}}
        \label{fig: sample 1 trajectory}
        \vspace{0.1cm}
        \includegraphics[width=\textwidth]{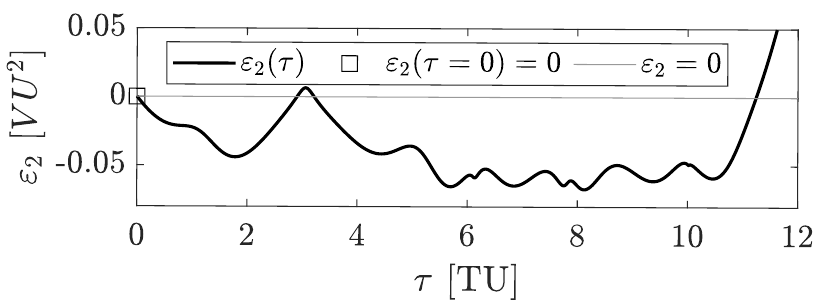}
        \subcaption{Two-body energy $\varepsilon_2$ over sample \gls{bc}~$\#1$}
    \end{subfigure}
    \hspace{0.8cm}
    \begin{subfigure}[t!]{0.43\textwidth}
        \includegraphics[width=\textwidth]{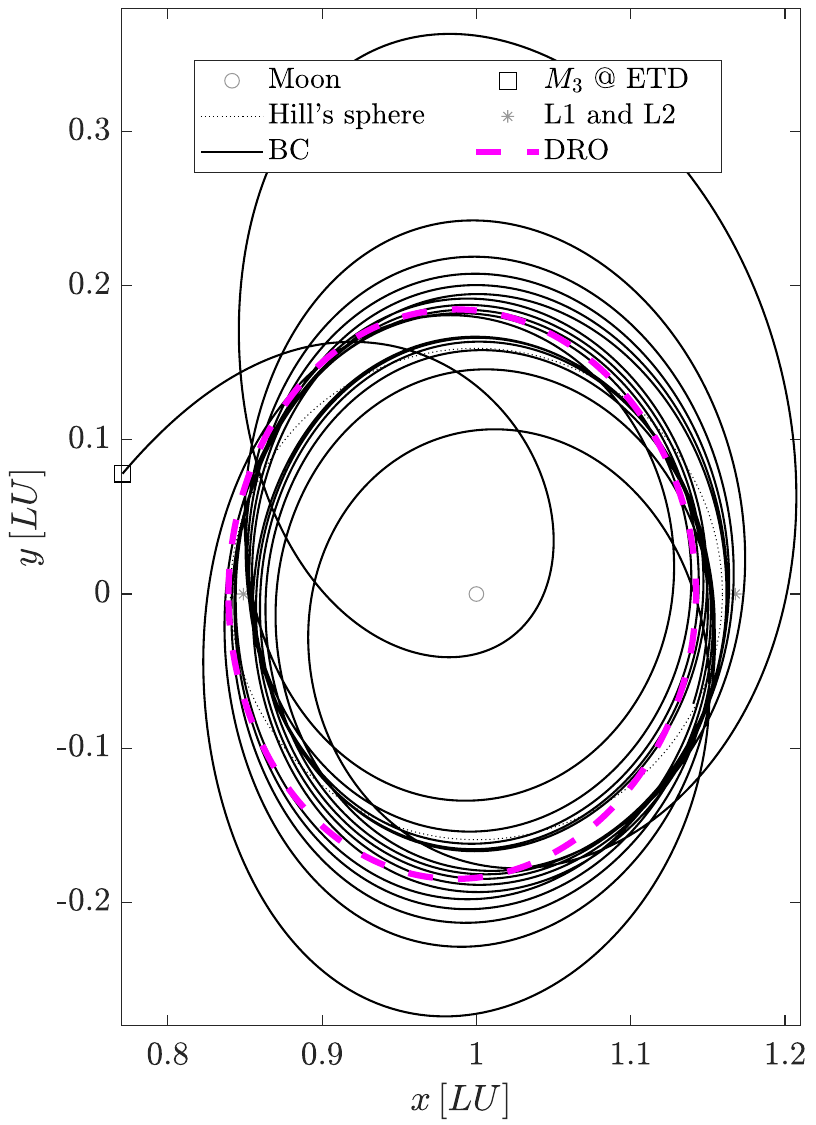}
        \subcaption{Sample \gls{bc}~$\#2$ trajectory from $\mathcal{C}(\Gamma = 1.18)$}
        \label{fig: sample 2 trajectory}
    \end{subfigure}
    \caption{Sample \gls{bc} trajectories $\#1$ and $\#2$, with driving \glspl{po} depicted with colored dashed lines.}
    \label{fig: sample 1 and 2 trajectories}
\end{figure}

With the sole intention of introducing the reader to the problem setup, two sample \gls{bc} trajectories are shown in \cref{fig: sample 1 and 2 trajectories}. Specifically, \cref{fig: sample 1 trajectory} corresponds to sample \gls{bc}~$\#1$, labeled in purple in \cref{fig: sample BCs on capture set}. This trajectory initially follows a path resembling the dynamics of a Lyapunov $L1$ \gls{po}, later transitioning to a \gls{dro}-like motion, thus exhibiting a distinct two-phase capture behavior. Conversely, \cref{fig: sample 2 trajectory} represents sample \gls{bc}~$\#2$, which belongs to $\mathcal{C}(\Gamma = 1.18)$ and therefore has a higher three-body energy compared to sample~$\#1$. It directly inserts into a retrograde motion around the Moon, exhibiting longer capture duration, as typically observed for retrograde captures at these energy levels~\cite{BC-ETDjournal}.

\begin{figure}[tb]
    \centering
    \includegraphics[width=0.42\textwidth]{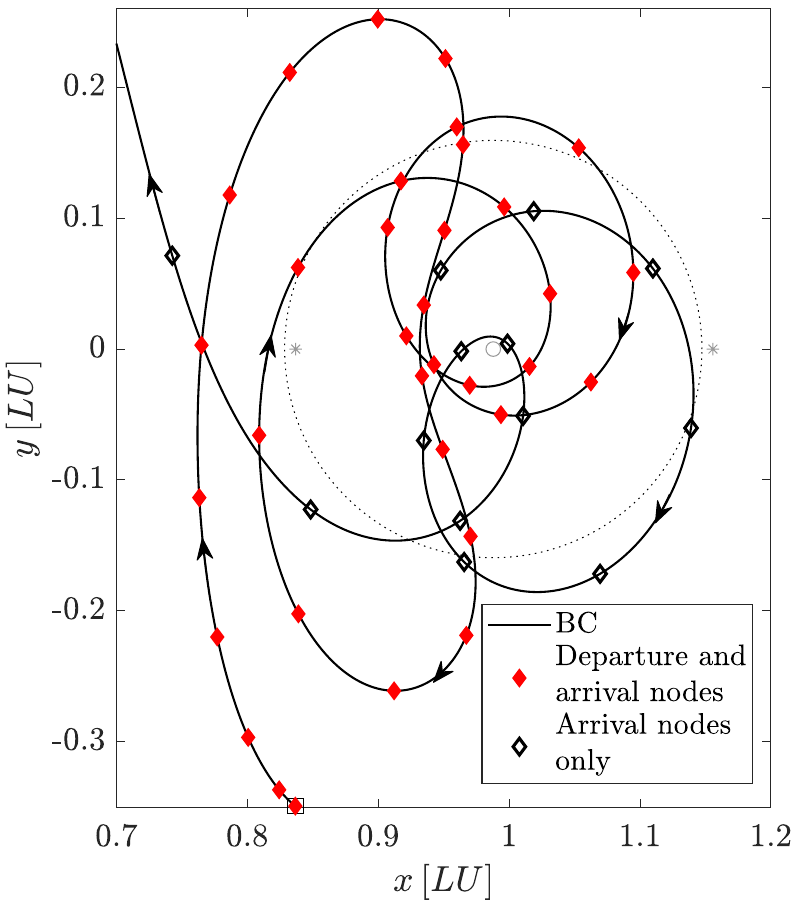}
    \caption{Sample \gls{bc}~$\#1$ nodes discretization (see \cref{fig: sample 1 trajectory}).}
    \label{fig: sample 1 nodes}
\end{figure}

Representative nodes are chosen along each \gls{bc} trajectory with an average timestep of $1$ day. Oversampling is applied near the beginning of the \gls{bc} and in regions where the distance to the Moon $r_2$ is smaller - phases that are typically more favorable in terms of transfer duration and cost, respectively. Additionally, when the spacecraft is close to the Moon, it traverses a larger arc in one day compared to when it is farther away, reinforcing the need for finer sampling in these segments. As a result, the sampling interval exceeds one day in the later phases of the \gls{bc} and/or when the spacecraft is farther from the Moon.
An example of these nodes, including arrows that indicate the direction of motion, is shown in \cref{fig: sample 1 nodes} for sample \gls{bc}~$\#1$ from \cref{fig: sample 1 trajectory}. All $n$ selected nodes are represented with diamond markers and serve dual purposes: as candidate locations for mono-impulsive insertion into a \gls{po}, and as initial guesses for the arrival nodes in bi-impulsive transfers.
Red-filled diamonds identify the subset of $n_0$ nodes occurring within the first $70\%$ of the total capture time, representing both departure and arrival nodes for the bi-impulsive transfers discussed in the following. In contrast, black diamonds represent arrival-only nodes located in the final $30\%$ of the \gls{bc}.
The choice for this restriction is twofold. First, it limits the optimization to transfers with shorter total durations, reducing computational effort and avoiding longer options that are less likely to comply with mission or timing constraints. Second, as discussed in~\cref{sec: BCs and POs poincare}, the last portion of the \gls{bc} before escape typically drifts away from nearby \glspl{po}, reducing its suitability for effective transfers. 

When introducing the transfer optimization method, we will use the index $i=1,2,\dots,n_0$ to indicate the departure nodes and $j=i+1,i+2,\dots,n$ for arrival nodes. The variable $\psi$ will be used to indicate the phase along the \gls{bc}, starting with $\psi_0=0$ for the first node where $i=0$, which corresponds to a state in the \gls{etd}. Finally, the dimensionless time after the \gls{etd} reads $\tau=\psi/(2\pi)$ and it is measured in TU (see \cref{tab: scaling units}).


\section{Connecting BCs and POs} \label{sec: relationship BCs and POs and mono-imp}
As established in the literature, there is a strong connection between \glspl{bc}, \glspl{po}~\cite{DeiTos-Russell-Topputo} and their associated manifolds~\cite{Topputo-Vasile-Bernelli}. Here, we investigate this connection by employing Poincaré section analysis to reduce the dimensionality of the planar problem and gain a deeper insight into the system's dynamics and structure. Building on this, the trade-off between transfer cost and total time is examined, and a method to compute mono-impulsive transfer costs from each \gls{bc} state to a specific \gls{po} family is presented.

As previously mentioned, the present analysis is limited to the planar problem. This restriction enables a clearer interpretation of the underlying dynamics and provides insights that are directly leveraged to initialize the planar optimization framework introduced in \cref{sec: opt section}. The extension to the spatial case is then developed in a subsequent section, following this foundational analysis.

\subsection{Poincaré section analysis} \label{sec: BCs and POs poincare}
\begin{figure}[tb]
    \begin{subfigure}[t!]{0.46\textwidth}
        \centering
        \includegraphics[width=\textwidth]{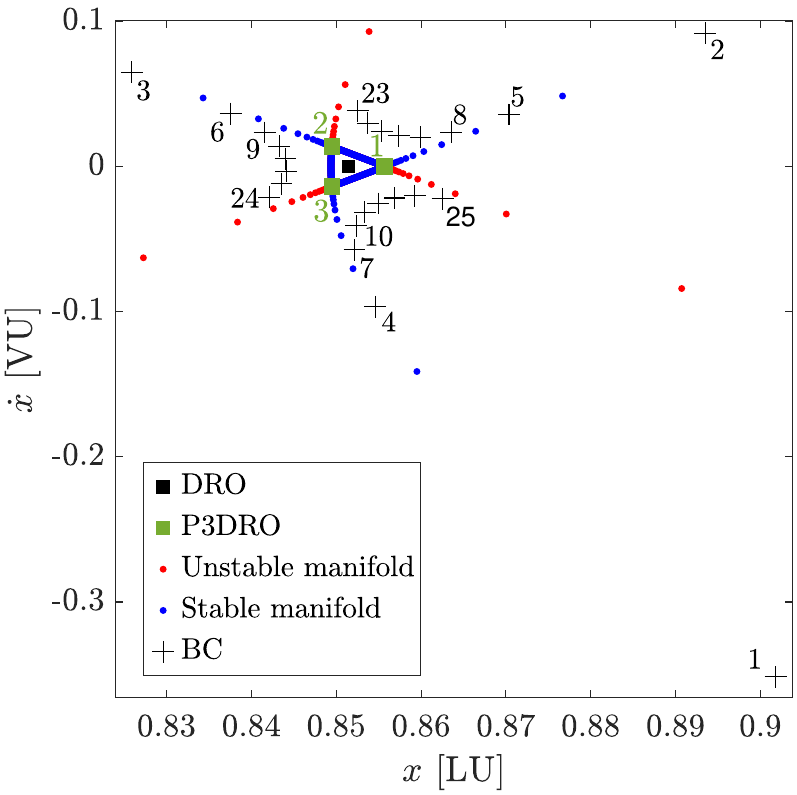}
        \subcaption{Sample \gls{bc}~$\#2$ and \gls{dro} families crossings in the Poincaré section}
        \label{fig: poincare map BC2}
    \end{subfigure} \hfill
    \begin{subfigure}[t!]{0.48\textwidth}
        \includegraphics[width=\textwidth]{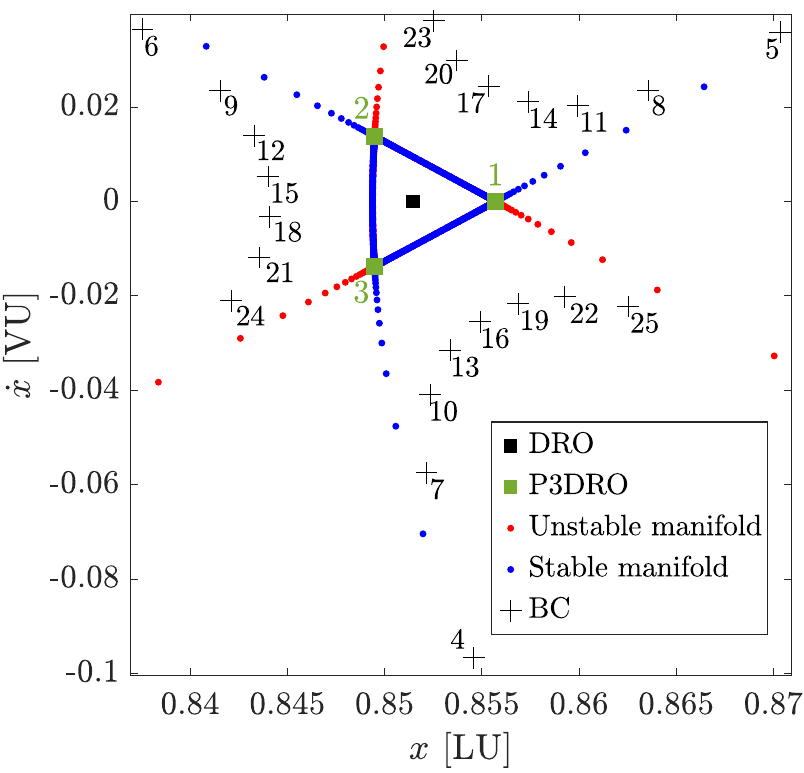}
        \subcaption{Close-up of \cref{fig: poincare map BC2}}
        \label{fig: poincare map BC2 ZOOM}
    \end{subfigure}
    \caption{Sample \gls{bc}~$\#2$ from \cref{fig: sample 2 trajectory} crossings on the Poincaré section (black plus signs) numbered in order of occurrence.}
    \label{fig: poincare map g3 and BC}
\end{figure}

Capdevila \textit{et al.}~\cite{Capdevila-Guzzetti-Howell2014} presented a particularly insightful Poincaré section representation of the \gls{dro} stability region, which is adapted and reproduced in \cref{fig: poincare map g3 and BC}. The Poincaré section is defined at $y=0$ in the synodic frame of the planar \gls{cr3bp}. Given that the Jacobi constant $C_J$ is conserved along a trajectory and can be used to compute $\dot{y}$, the four-dimensional state space reduces to a two-dimensional map on $x$ and $\dot{x}$. The remaining two coordinates are $y=0$ and $\dot{y}=f(C_J,x,\dot{x})$. The points of two different \glspl{po} crossing this section are represented in black and green, and they respectively belong to the \gls{dro} family f and \gls{dro} family g3~\cite{henon1969V} (period-tripled also known as f3~\cite{aydin2025POfamilies}, bifurcating from the family f~\cite{broucke1969stability, Markellos1974}). In the present work, the \gls{dro} family g3 will be addressed as \gls{p3dro}. In addition, the blue/red dotted lines represent the stable/unstable manifold maps of the \gls{p3dro}, as they emanate to/from the green points of intersection with the selected Poincaré section. The triangular region enclosed by the green vertices is known as \gls{dro} stability region, where \glspl{qpo} (non-periodic stable orbits) can be found.
The intersections of sample \gls{bc}~$\#2$ 
with the same Poincaré section are mapped with black plus signs and are numbered in time order.

As expected from dynamical systems theory, the invariant manifolds associated with \glspl{po} strongly influence the behavior of nearby trajectories. This is clearly illustrated in~\cref{fig: poincare map g3 and BC}, where the dynamics of this particular \gls{bc} are governed by the manifolds of the \gls{p3dro}. More generally, the results presented in the following sections indicate that, for any of the considered \gls{bc} trajectories, one or more \glspl{po} can be identified that explain their motion and share similar dynamical characteristics.
The driving \gls{po} typically varies with the Jacobi constant (or, equivalently, the three-body energy parameter $\Gamma$), and may also change over the course of a trajectory, as different phases of a \gls{bc} can be governed by different families' dynamics.
An example is provided by sample \gls{bc}~$\#1$ in \cref{fig: sample 1 trajectory}, which is at first strongly influenced by the Lyapunov $L1$ and then by the \gls{dro}/\gls{p3dro} dynamics.
The remainder of this work presents results and targeted analyses that support and clarify these initial hypotheses and visual observations.

\subsection{Mono-impulsive cost estimate to iso-energy PO} \label{sec: mono-imp cost estimate poincare}

A preliminary discussion on the transfer cost from a \gls{bc} to an iso-energy \gls{po} can be based simply on \cref{fig: poincare map g3 and BC}. Exploiting the Poincaré representation, the insertion cost into the \gls{dro} stability region is estimated as a function of the current \gls{bc} phase $\psi$. As previously mentioned, the latter is related to the dimensionless time $\tau=\psi/(2\pi)$; therefore, the waiting time $t_{wait}$ in days is given by $t_{wait}=\tau \cdot TU$. In fact, the relative distance between the \gls{bc} intersection points and the \gls{dro} stability region contained within the green \gls{p3dro} points is used to characterize a single impulse transfer from the \gls{bc} to a \gls{po} or \gls{qpo}. The cost usually decreases over time, revealing a trade-off: a shorter wait implies a higher $\Delta v$, while a longer wait can significantly reduce the injection cost. In this example, the minimum mono-impulsive injection occurs at the 16\textsuperscript{th} intersection with the Poincaré section, implying a required wait of approximately $t_{wait} \approx 180$ days. This indeed represents a significant delay, even though a hypothetical mission would spend this time in a weakly unstable orbit near (and asymptotically approaching) the target \gls{po}. The maneuver cost is estimated by measuring the correction in the $\dot{x}$ component and adding the correction in $\dot{y}=f(C_J,x,\dot{x})$. For the 16\textsuperscript{th} intersection with the Poincaré section in \cref{fig: poincare map g3 and BC}, the cost is only $\Delta v_{min} \approx 26$ m/s.

\subsection{Mono-impulsive cost to PO families using the abacus} \label{sec: mono-imp cost to family}
When considering \gls{dro} or Lyapunov families, there is a one-to-one correspondence between a planar \gls{bc} position $(x_{BC},y_{BC})$ and some parameters $(p,\varphi)$ that defines $(x, y)_{PO}$ on a \gls{po} within the chosen family. In general, every position state in the cislunar region is linked to one and only one position state of the \gls{dro} or Lyapunov families. As a consequence, from each position, a velocity correction is computed to achieve a mono-impulsive insertion into a \gls{po}.

Using the abacus introduced in \cref{sec: abacus of POs}, this correspondence in the position state is achieved via an iterative search over an adaptive grid. A selected family within the abacus is evaluated initially over its entire domain. Then, the grid is progressively refined until a pair $(p_f, \varphi_f)$ is found such that $\sqrt{(x_{PO} - x_{\text{BC}})^2 + (y_{PO} - y_{\text{BC}})^2} < \epsilon$, where $\epsilon = 10^{-8}$.
Thanks to the parameters $(p_f, \varphi_f)$, the entire state $\mathbf{x}_{PO}(p_f, \varphi_f)$ is retrieved, and a mono-impulsive correction for a transfer from $\mathbf{x}_{BC}$ to $\mathbf{x}_{PO}$ is computed as
\begin{equation}
    \Delta v_{\text{mono}}=\sqrt{(\dot{x}_{PO}-\dot{x}_{BC})^2+(\dot{y}_{PO}-\dot{y}_{BC})^2} \, .
    \label{eq: deltav mono}
\end{equation}
Note that for planar \glspl{bc}, and for both \gls{dro} and Lyapunov families, $z$ and $\dot{z}$ components are always null. For this reason, they are not introduced here. Instead, the spatial case will be discussed in \cref{sec: spatial problem}.
This maneuver cost $\Delta v_{\text{mono}}$ is computed for all $n$ nodes marked with black triangles in \cref{fig: sample 1 nodes}, each representing a mono-impulsive transfer option to a selected \gls{po} family. These same transfers also serve as initial guesses for the bi-impulsive optimization method introduced in the next section, where the associated mono- vs bi-impulsive costs are examined.

\section{Optimization of bi-impulsive transfers} \label{sec: opt section}

In this section, the cost-estimation method previously introduced is extended to initialize the planar bi-impulsive transfer optimization process. The core steps are described in \cref{sec: core steps}, while Sections~\ref{sec: follow minimum} and \ref{sec: span on phase} complement these steps to ensure comprehensive coverage of the optimization variable space.
The formulation proposed for the optimization is first applied to the planar problem, as detailed in the following; the same optimization procedure (excluding the seeding strategy) is later employed without modification for the spatial case in \cref{sec: spatial problem}.

To define a transfer trajectory between a \gls{bc} and a \gls{po}, four design variables are used: an initial phase $\psi$ from the departure \gls{bc}, a final target phase $\varphi$ on the target \gls{po}, the family parameter $p$,
and the \gls{tof}. 
The method presented here finds an optimal transfer trajectory starting from a fixed phase $\psi_i$. 
Instead, $p$, $\varphi$, and $\text{ToF}$ are the optimization variables, whose local optimum is indicated by $(p^*, \varphi^*, \text{ToF}^*)$. Although $\psi_i$ is fixed for each individual optimization, multiple values are considered across a discrete sweep from $\psi_i = \psi_0 = 0$ to $\psi_i = \psi_{n_0}$ (see \cref{sec: BC introduction} and nodes of \cref{fig: sample 1 nodes}). 
In this way, the dependence on every possible variable of the bi-impulsive transfer problem is investigated. Nevertheless, the transfer optimality is limited in this variable, as $\psi_i$ is treated as a discretized parameter rather than a continuously optimized free variable.
Even though the optimization framework could easily accommodate an additional variable, it was excluded to reduce computational cost and avoid unnecessary complexity.

The fixed initial phase on the \gls{bc} and the target point on the \gls{po} 
are respectively expressed as:
\begin{equation}
      \mathbf{x}_0 = \mathbf{x}_{BC} \left(\psi_i \right) \, , \qquad \mathbf{x}_f = \mathbf{x}_{PO} \left(p,\varphi \right) \, .
      \label{eq: initial and final states}
\end{equation}
Each point $\mathbf{x}_f$ has a corresponding period, which is called $T(p)$ and does not depend on the phase $\varphi$.
In addition, $\mathbf{x}_{BC,f}$ denotes the final state obtained by propagating the initial condition $\mathbf{x}_0$ forward for a duration of $\text{ToF}$.

\begin{figure}[tb]
    \centering
    \includegraphics[width=0.42\textwidth]{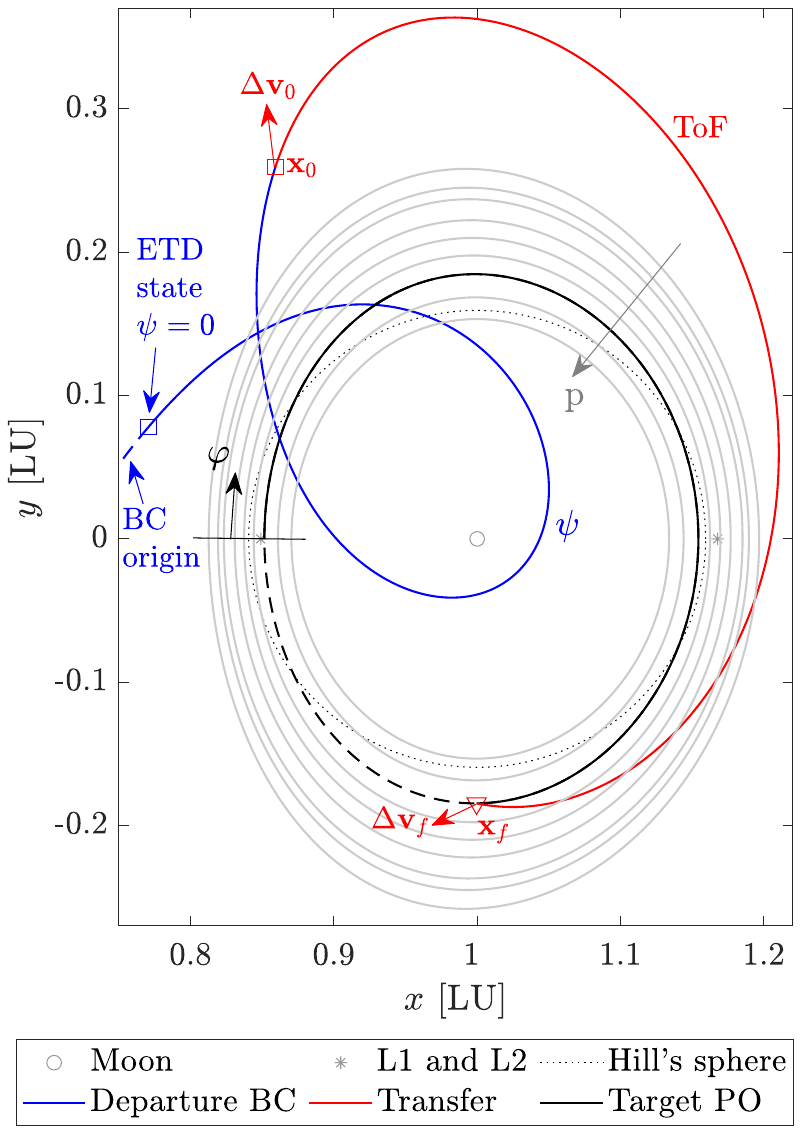}
    \caption{Sketch of a bi-impulsive transfer from \gls{bc} to the \gls{dro} family.}
    \label{fig: sketch transfer}
\end{figure}

The optimization problem is illustrated in \cref{fig: sketch transfer}.
To ensure coverage of the entire span of the target phase $\varphi$, the algorithm performs multiple independent optimization procedures, each initialized with a pair of indices $i$ and $j$ (see the end of \cref{sec: BC introduction}) that create an exhaustive combination of transfers between each departure and arrival node. As a consequence, in each independent optimization step $j$, the variable $\varphi$ is bounded within the interval $\varphi_{j-1}<\varphi<\varphi_{j+1}$.

The following subsections describe the initialization strategy, followed by the core optimization step, and finally the overall enclosing algorithm for the bi-impulsive transfer optimization method.

\subsection{Seeding the bi-impulsive optimization with mono-impulsive solutions} \label{sec: initial guess}
The mono-impulsive transfers computed in \cref{sec: mono-imp cost to family} are repurposed here to initialize the search for bi-impulsive solutions.
Specifically, each mono-impulsive solution defines the target endpoint of a bi-impulsive transfer, where the final maneuver $\Delta v_f=\Delta v_{\text{mono}}$ must insert the spacecraft into the same \gls{po}. The initial maneuver $\Delta v_0$ is set to zero at first, effectively leveraging the natural propagation along the \gls{bc} up to the node $j$ where $\Delta v_f$ is applied, providing a simple yet informed starting point for the optimization. 
In addition, the \gls{po} parameters describing the target state $\mathbf{x}_f$ are initialized as $(p, \varphi)=(p_f, \varphi_f)$ (see \cref{sec: mono-imp cost to family}). The index $k$ of the $k$-th polynomial map in the abacus $\mathcal{M}^k_{PO}$ describing the neighborhood in the $(p_f, \varphi_f)$ space is also extracted for later use (see \cref{sec: abacus of POs}).
In the same fashion, the \gls{tof} is initialized as $\text{ToF}=\tau_f-\tau_0=\tau(\psi_j)-\tau(\psi_i)$, where $\psi_j$ is the phase of the current $j$-th \gls{bc} node considered, i.e. when the second maneuver $\Delta v_f$ is applied.

\subsection{Core of the optimization procedure} \label{sec: core steps}

The first key step consists of expanding the dynamics around the reference \gls{bc} trajectory between any two nodes over a Time of Flight $\text{ToF} = \tau_f - \tau_0$. With the \gls{da} polynomial order set to 8, a high-order expansion propagates the state from the initial condition $\mathbf{x}_0 = [ \mathbf{r}_0 ; \mathbf{v}_0 ]$ to the final state $\mathbf{x}_f = [ \mathbf{r}_f ; \mathbf{v}_f ]$.
A polynomial representation of the time dependence on the final state is introduced through a \gls{da} variable $\delta \text{ToF}$.
Therefore, the equations of motion $\dot{\mathbf{x}} = \mathbf{f}(\mathbf{x}, \tau)$ in \cref{eq: equations of motion} are reformulated as:
\begin{equation}
    \begin{cases}
        \cfrac{d\mathbf{x}}{d\chi} = \text{ToF} \cdot \mathbf{f}(\mathbf{x},\tau) \\[5pt]
        \cfrac{d(\text{ToF})}{d\chi} = 0 \\[5pt]
    \end{cases} \, .
    \label{eq: dynamics with variable time}
\end{equation}
Here, $\chi \in [0, 1]$ is an artificial and independent propagation variable used solely for the expansion with respect to $\delta \text{ToF}$, which represents physical time through $\tau=\text{ToF} \cdot \chi + \tau_0$.

In addition, \gls{da} is used to map the influence of an initial correction $\delta \mathbf{v}_0$ applied to the initial velocity $\mathbf{v}_0$ in Cartesian coordinates.
To quantify how variations in initial conditions affect the final state, we compute a \gls{da}-based polynomial map of the propagated dynamics:
\begin{equation}
    \begin{pmatrix}
        \delta \mathbf{r}_f\\[\jot]
        \delta \mathbf{v}_f\\[\jot]
        \delta \text{ToF}
    \end{pmatrix}
    =
    \begin{pmatrix}
        \mathcal{M}_{\mathbf{r}_f} \\[\jot]
        \mathcal{M}_{\mathbf{v}_f} \\[\jot]
        \mathcal{I}
    \end{pmatrix}
    \begin{pmatrix}
        \delta \mathbf{v}_0 \\[\jot]
        \delta \text{ToF}
    \end{pmatrix} \, ,
    \label{eq: MAPD}
\end{equation}
where $\mathcal{I}$ represents the identity function.
Then, we consider the sub-map:
\begin{equation}
    \begin{pmatrix}
        \delta \mathbf{r}_f\\[\jot]
        \delta \text{ToF}
    \end{pmatrix}
    =
    \begin{pmatrix}
        \mathcal{M}_{\mathbf{r}_f} \\[\jot]
        \mathcal{I}
    \end{pmatrix}
    \begin{pmatrix}
        \delta \mathbf{v}_0 \\[\jot]
        \delta \text{ToF}
    \end{pmatrix} \, ,
    \label{eq: MAPD2}
\end{equation}
which maps three input variables to three outputs. To solve the \gls{tpbvp} using \gls{da}, this map is inverted~\cite{berz1992} using polynomial inversion techniques to obtain:
\begin{equation}
    \begin{pmatrix}
        \delta \mathbf{v}_0 \\[\jot]
        \delta \text{ToF}
    \end{pmatrix}
    =
    \begin{pmatrix}
        \mathcal{M}_{\mathbf{r}_f} \\[\jot]
        \mathcal{I}
    \end{pmatrix}
    ^{-1}
    \begin{pmatrix}
        \delta \mathbf{r}_f\\[\jot]
        \delta \text{ToF}
    \end{pmatrix} \, .
    \label{eq: MAPI}
\end{equation}
This inverted map represents a polynomial expansion of the dynamics in the neighborhood of the reference trajectory: it provides the required corrections $\delta \mathbf{v}_0$ and $\delta \text{ToF}$ to the initial velocity and propagation time needed to reach a perturbed final position $\delta \mathbf{r}_f$~\cite{lizia2008application}.
Unlike classical point-wise shooting methods~\cite{stoer1980introduction}, which require iterative integration, this formulation provides a continuous representation of the \gls{tpbvp} solution and enables the evaluation of multiple trajectory corrections from a single propagation~\cite{armellin2006TPBVP}. 

At this point, the local map $\mathcal{M}^k_{PO}$ of the target \gls{po} family, introduced in \cref{sec: initial guess}, is used to set up the \gls{tpbvp}. The target points around the nominal trajectory are described as a function of the \gls{po} parameters $(p, \varphi)$:
\begin{equation}
    \delta \mathbf{r}_f = \delta \mathbf{r}_f(\delta p, \delta \varphi) \, .
    \label{eq: deltarf function of p phi}
\end{equation}
Therefore, the composition of \cref{eq: MAPI} with \cref{eq: deltarf function of p phi} allows for the computation of the first maneuver map:
\begin{equation}
    \delta \mathbf{v}_0 = \mathcal{M}_{\delta \mathbf{v}_0} \left( \delta p, \delta \varphi, \delta \text{ToF} \right) \, ,
    \label{eq: deltaV0 map}
\end{equation}
where the two additional \gls{da} variables $(\delta p, \delta \varphi)$ represent the perturbation around the nominal values $(p_f, \varphi_f)$. This vector of polynomial maps 
approximates the initial impulse $\Delta \mathbf{v}_0 = \delta \mathbf{v}_0$ required to reach a \gls{po} within the target family as a function of the \gls{po} parameters and \gls{tof}.
The composition of the velocity map $\mathcal{M}_{\mathbf{v}_f}(\delta \mathbf{v}_0,\delta \text{ToF})$ in \cref{eq: MAPD} with $\delta \mathbf{v}_0$, returns a fully parametric expression for $\delta \mathbf{v}_f = \mathcal{M}_{\delta \mathbf{v}_f}\left( \delta p, \delta \varphi, \delta \text{ToF} \right)$. The arrival impulse is then readily obtained by 
\begin{equation}
    \Delta \mathbf{v}_f = \mathbf{v}_{PO,f}\left( \delta p, \delta \varphi \right) - \left( \mathbf{v}_{BC,f} + \mathcal{M}_{\delta \mathbf{v}_f}\left( \delta p, \delta \varphi, \delta \text{ToF} \right)\right)
    \label{eq: Deltav1 map}
\end{equation}
where $\mathbf{v}_{PO,f}$ is again extracted from the local map $\mathcal{M}^k_{PO}$ of the target \gls{po} family.

A convergence radius $\rho_{\text{ToF}}$ for the maps $\Delta \mathbf{v}_0$ and $\Delta \mathbf{v}_f$ is estimated in terms of $\delta \text{ToF}$. In this work, a tolerance of approximately $10^{-3}$ m/s is employed to estimate the convergence radii. This means that the accuracy of the map is not guaranteed outside the range 
\begin{equation}
    \delta \text{ToF} \in I_t = [I_{t,l}, \, I_{t,u}] = [-\rho_{\text{ToF}}, \, +\rho_{\text{ToF}}] \, ,
    \label{eq: delta t interval}
\end{equation}
where a new polynomial expansion of the dynamics is required with a refined guess for the $\text{ToF}$ variable.

The same reasoning applies to $\delta p$ and $\delta \varphi$.
To monitor the accuracy of the map $\Delta \mathbf{v}_0$, a convergence radius $\rho_{rf}$ is estimated in terms of $\delta \mathbf{r}_f$, with a tolerance threshold of $10^{-7}$ LU (approximately 40 m). The expansion in \cref{eq: MAPI} is considered valid as long as $\|\delta \mathbf{r}_f\| < \rho_{rf}$.
In contrast, the accuracy of the map $\Delta \mathbf{v}_f$ is assessed only a posteriori, as it does not affect the feasibility of the transfer but impacts only the precision of the cost estimate.
Finally, the validity interval of the current \gls{po} family map $\mathcal{M}^k_{PO}$ must be enforced whenever it imposes a tighter constraint than $\rho_{rf}$.
As a consequence, the optimization range in $\delta p$ reads
\begin{equation}
    \delta p \in I_p = [I_{p,l}, \, I_{p,u}] = [max(-\rho_{rf},p_{k,l}-p_f), \, min(+\rho_{rf},p_{k,u}-p_f)] \, ,
    \label{eq: delta p interval}
\end{equation}
where $p_{k,l}$ and $p_{k,u}$ are boundary values of the current $k$-th map $\mathcal{M}^k_{PO}$ of the abacus, as introduced in \cref{sec: abacus of POs}. From \cref{sec: mono-imp cost to family}, $p_f$ is initially obtained and then updated during the optimization algorithm introduced in the following. Note that the quantities $\rho_{rf}$ and $p$ are both measured in LU, and hence can be directly compared.
A similar procedure takes place for the phase $\varphi$:
\begin{equation}
    \delta \varphi \in I_\varphi = [I_{\varphi,l}, \, I_{\varphi,u}] = [max(-\rho_{rf}/r_2,\varphi_{k,l}-\varphi_f), \, min(+\rho_{rf}/r_2,\varphi_{k,u}-\varphi_f)]
    \label{eq: delta phi interval}
\end{equation}
where the only difference lies in the presence of the denominator $r_2$, as introduced in \cref{eq: r_2}. This is introduced to ensure dimensional consistency when comparing $\rho_{rf}$ with phase variations. The actual value for $r_2$ is computed for the nominal trajectory and considered uniform for the entire polynomial expansion of the final state $\mathbf{x}_f$.
In the following, $\mathring{I}_p=\,]I_{p,l}, \, I_{p,u}[$, $\mathring{I}_\varphi=\,]I_{\varphi,l}, \, I_{\varphi,u}[$, and $\mathring{I}_t=\,]I_{t,l}, \, I_{t,u}[$ will be used to denote the interior set of the interval, hence excluding the boundaries of the intervals.


Finally, a function describing the total $\Delta v$ for the transfer trajectory solving the \gls{tpbvp} and determining the optimal insertion into the target \gls{po} family is the cost function
\begin{equation}
    J\left(\delta p,\delta \varphi, \delta ToF\right) = \Delta v = \Delta v_0 + \Delta v_f \ = \mathcal{M}_{\Delta v}(\delta p, \delta \varphi, \delta \text{ToF}) \, ,
    \label{eq: cost function}
\end{equation}
where $\Delta v_0$ and $\Delta v_f$ are obtained by applying the Euclidean norm function to the initial and final velocity correction maps $\Delta \mathbf{v}_0$ and $\Delta \mathbf{v}_f$, respectively.
As a consequence, the map $\mathcal{M}_{\Delta v}(\delta p, \delta \varphi, \delta \text{ToF})$ enables the computation of the optimal values of the differential variables $(\delta p^*, \delta \varphi^*, \delta \text{ToF}^*)$ that minimize the total impulse $\Delta v^*$.
These variables are defined with respect to the nominal parameters $(p_f, \varphi_f, \text{ToF})$, from which the actual optimal parameters are recovered as $p^* = p_f + \delta p^*$, $\varphi^* = \varphi_f + \delta \varphi^*$, and $\text{ToF}^* = \text{ToF} + \delta \text{ToF}^*$.
However, because the norm operator introduces nonlinearities (particularly due to the square root), it cannot be directly applied within the component-wise polynomial map framework without degrading accuracy. For this reason, the cost function in \cref{eq: cost function} must be reconstructed internally by the optimizer from its individual components. As such, the symbolic maps provided as input must separately represent each component: $\Delta v_{0,x}$, $\Delta v_{0,y}$, $\Delta v_{0,z}$, $\Delta v_{f,x}$, $\Delta v_{f,y}$, and $\Delta v_{f,z}$.

The BFGS quasi-newton method implemented in the \textit{find\textunderscore min\textunderscore box\textunderscore constrained} general-purpose non-linear optimizer of the DLIB library\footnote{Library available at: \url{https://dlib.net/} [last accessed \lastdate].}~\cite{dlib} is used in this work. This optimizer takes as input the cost function $J\left(\delta p,\delta \varphi, \delta ToF\right)$ itself, as well as its derivative with respect to the optimization variables $\delta p$, $\delta \varphi$, and $\delta ToF$. Having already computed the polynomial maps, these derivatives are included in the available expansions; therefore, the gradient is extracted with no further computations. In addition, the validity intervals $I_p$, $I_\varphi$, and $I_t$ define a constraint box in the 3D search space, which is passed to the DLIB solver. To improve the likelihood of identifying the overall minimum within the search domain, the optimizer is initialized from multiple starting points. Specifically, nine initial guesses are used: the expansion point $\left( \delta p=0, \delta \varphi=0, \delta \text{ToF}=0 \right)$, along with the eight corners of the box.
Each starting point is independently passed to the DLIB solver, and the resulting solutions are compared. The transfer yielding the lowest cost $J$ is retained as the final optimized transfer.

\subsection{Following the local minimum} \label{sec: follow minimum}
The optimization method presented in \cref{sec: core steps} computes the local minimum within the boundaries of the box.
If the local minimum is located within the box, the procedure is stopped, and the parameters describing the minimum cost are stored.
However, in some cases, the DLIB optimizer returns a solution located on the boundary of the search box. This indicates that the minimum for the cost function could be located outside the current bounds, but the optimizer cannot reach it.
To address this event, the optimization process of \cref{sec: core steps} is encapsulated in an algorithm that adaptively follows the minimum of the cost function by performing a new expansion of the dynamics and, if needed, selects a different local \gls{po} family map $\mathcal{M}^k_{PO}$ by adjusting the index $k$.
This process is summarized in \cref{alg: algorithm follow minimum}, where the boundary‐hit logic is grouped into three helper steps:

\begin{itemize}
  \item Convergence check (lines \ref{linealg: Check Start}-\ref{linealg: Check End}): checks accuracy of the solution $(\delta p^*, \delta\varphi^*, \delta\mathrm{ToF}^*)$ obtained.
  \item Accept solution (lines \ref{linealg: AcceptSol Start}-\ref{linealg: AcceptSol End}): saves the current best solution when this is found strictly inside the box or when phases of adjacent nodes are reached. 
  \item Handle boundary (lines \ref{linealg: HandleBoundary Start}-\ref{linealg: HandleBoundary End}): when the boundary of the box is reached (parameter, phase, and/or time), appropriate map update(s) are applied.
\end{itemize}
A maximum of iterations $\text{iter}_{max}=20$ is employed to avoid excessive computational cost, particularly when extremely small convergence radius values cause slow progress. In some cases, non-convergence within this limit of iterations or other numerical issues arise due to the intentionally large variable space; these are typically observed after reaching a local minimum and are robustly handled to prevent the failure of the entire process, allowing the optimizer to continue exploring the remaining regions of the variable space.

\begin{algorithm}[tb]
\caption{Core enveloping algorithm: following local minimum.}
\label{alg: algorithm follow minimum}
\begin{algorithmic}[1]
\State{Given an initial guess from \cref{sec: mono-imp cost to family} connecting departure node $i$ to arrival node $j$, set $\mathbf{x}_0=\mathbf{x}_{BC}(\psi_i)$.} 
\State{Set nominal (expansion) parameters $p_f=p_j$, $\varphi_f=\varphi_j$ and $\text{ToF}=t(\psi_j)-t(\psi_i)$. Set $\text{iter}=0$ and $\text{iter}_{max}=20$.} 
\While{$\text{iter} < \text{iter}_{max}$}  \label{linealg: start while alg core} \Comment{The optimization variables $\delta p$, $\delta \varphi$, and $\delta ToF$ are updated until a local minimum is reached.}
    \State{$\text{iter} \gets \text{iter}+1$}
    \State{Follow procedure of \cref{sec: core steps}, obtaining minimum $\Delta v^*$ in box for the variables $\delta p^*$, $\delta \varphi^*$, and $\delta \text{ToF}^*$.}
    \State{$\text{ePos}= \mathbf{r}_{PO,f}(p^*,\varphi^*)-[\mathbf{r}_{BC,f}+\delta \mathbf{r}_f(\delta p^*, \delta \varphi^*, \delta \text{ToF}^*)]$} \Comment{Compute solution error in the position} \label{linealg: Check Start}
    \State{$\text{eVel}= \mathbf{v}_{PO,f}(p^*,\varphi^*)-[\mathbf{v}_{BC,f}+\delta \mathbf{v}_f(\delta p^*, \delta \varphi^*, \delta \text{ToF}^*)]$.} \Comment{Compute solution error in the velocity}
    \If{$\left(\text{ePos}>10^{-5} \, \text{LU} \approx 4 \, \text{km}\right) \; \lor \; \left(\text{eVel}>10^{-2} \, \text{VU} \approx 10 \, \text{m/s}\right)$} 
        \State{Retrieve solution from previous iteration and store its $\Delta v^*$, $p^*$, $\varphi^*$, $\text{ToF}^*$, $\Delta\mathbf{v}_0^*$, and $\Delta\mathbf{v}_f^*$. \underline{Exit while loop}.}
    \EndIf \label{linealg: Check End}
    
    \If{$\delta \text{ToF}^* \in \mathring{I}_t \land \delta p^* \in \mathring{I}_p \land \delta \varphi^* \in \mathring{I}_\varphi$} \Comment{The local minimum is found strictly inside the current box} \label{linealg: AcceptSol Start}
        \State{Store $\Delta v^*$, $p^*$, $\varphi^*$, $\text{ToF}^*$, $\Delta\mathbf{v}_0^*$, and $\Delta\mathbf{v}_f^*$. \underline{Exit while loop}.}
    \EndIf \label{linealg: AcceptSol End}
    
    \If{$\delta \varphi^* = I_{\varphi,l} \lor \delta \varphi^* = I_{\varphi,u}$} \Comment{The solution is in the phase-boundary of the validity box} \label{linealg: HandleBoundary Start}
        \If{$\delta \varphi^* = \varphi_{j-1}-\varphi \lor \delta \varphi^* = \varphi_{j+1}-\varphi$} \Comment{Boundary of phase span defined by neighboring nodes}
            \State{Store $\Delta v^*$, $p^*$, $\varphi^*$, $\text{ToF}^*$, $\Delta\mathbf{v}_0^*$, and $\Delta\mathbf{v}_f^*$. \underline{Exit while loop}.}
        \ElsIf{$\varphi^* = \varphi_{k,l} \lor \varphi^* = \varphi_{k,u}$} \Comment{Boundary of the current \gls{po} family map $\mathcal{M}^k_{PO}$ interval validity}
            \State{Force an update of the \gls{po} family map by setting $\delta \varphi^* = (1+\epsilon)\delta \varphi^*$, with a small $\epsilon$ (e.g. $\epsilon=10^{-5}$).}
        \EndIf
    \EndIf
    
    \If{$\delta p^* = I_{p,l} \lor \delta p^* = I_{p,u}$} \Comment{The solution is in the parameter-boundary of the validity box}
        \If{$p^* = p_{k,l} \lor p^* = p_{k,u}$} \Comment{Boundary of the current \gls{po} family map $\mathcal{M}^k_{PO}$ interval validity}
            \State{Force an update of the \gls{po} family map by setting $\delta p^* = (1+\epsilon)\delta p^*$, with a small $\epsilon$ (e.g. $\epsilon=10^{-5}$).}
        \EndIf
    \EndIf \label{linealg: HandleBoundary End}
    
    \State{Update $p_f = p_f + \delta p^*$, $\varphi_f = \varphi_f + \delta \varphi^*$, and $\text{ToF} = \text{ToF} + \delta \text{ToF}^*$.} 
    \EndWhile \label{linealg: end while alg core}
    \State{Store $\Delta v^*$, $p^*$, $\varphi^*$, $\text{ToF}^*$, $\Delta\mathbf{v}_0^*$, and $\Delta\mathbf{v}_f^*$.} \Comment{At $\text{iter}=\text{iter}_{max}$, accept current solution}
\end{algorithmic}
\end{algorithm}

Although more complex to implement, this approach avoids relying on \gls{ads}~\cite{wittig2015propagation} to construct an exhaustive domain $[\delta p,\delta \varphi, \delta ToF]$, which implies creating a very broad expansion domain in terms of $[\delta \mathbf{v}_0, \delta ToF]$.
Such an approach would be computationally expensive and inherently limited by the predefined expansion domain, potentially missing valid local minima lying outside it. Instead, the current method expands the dynamics locally and only where needed, allowing the optimizer to consistently follow the gradient toward a local minimum.

\subsection{Spanning along the PO phase} \label{sec: span on phase}
\begin{figure}[tb]
    \centering
    \begin{subfigure}[t!]{0.47\textwidth}
        \centering
        \includegraphics[width=\textwidth]{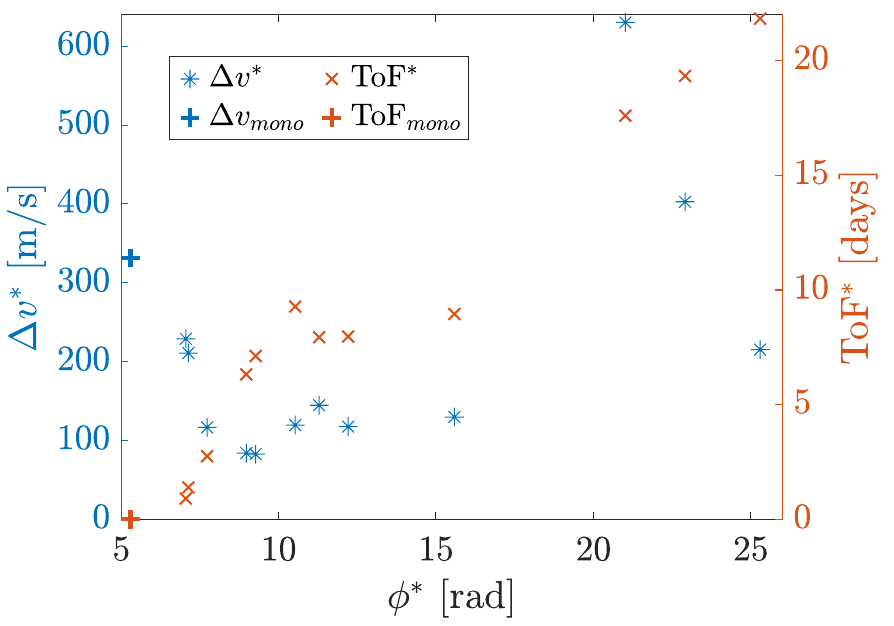}
        \subcaption{Transfer cost $\Delta v^*$ and $\text{ToF}^*$ against $\varphi^*$}
        \label{fig: optimal sols fixed psi DV and ToF vs phi}
    \end{subfigure} \hspace{0.7cm}
    \begin{subfigure}[t!]{0.43\textwidth}
        \includegraphics[width=\textwidth]{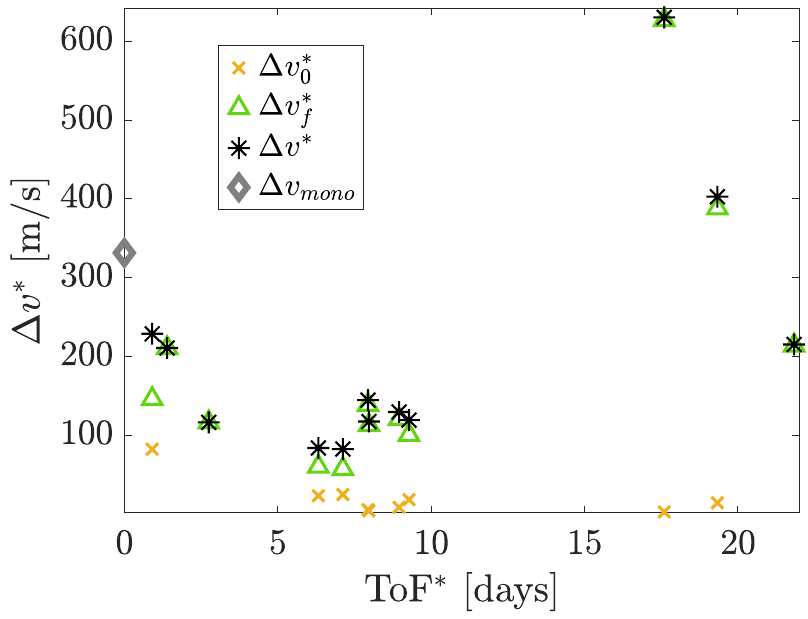}
        \subcaption{Transfer cost $\Delta v^*$ against $\text{ToF}^*$}
        \label{fig: optimal sols fixed psi DVs vs ToF}
    \end{subfigure}
    \caption{Optimal solutions from \cref{alg: algorithm span phase PO} for $\forall j \land i=26$ node of sample \gls{bc}~$\#1$  introduced in \cref{fig: sample 1 trajectory}.}
    \label{fig: optimal sols fixed psi}
\end{figure}

While keeping the initial state $\mathbf{x}_0$ fixed (i.e. fixed $\psi_i$), the algorithm presented in \cref{sec: follow minimum} is repeated for each node $j$, spanning on all the possible \gls{po} phases, while constraining $\varphi \in [\varphi_{j-1}, \varphi_{j+1}]$, as introduced in \cref{alg: algorithm span phase PO}.
The results obtained are represented in \cref{fig: optimal sols fixed psi}, where mono-impulsive solutions are also included and indicated with plus signs. In particular, the optimal cost $\Delta v^*$ and $\text{ToF}^*$ are represented against the spanned arrival phase on the \gls{po} family $\varphi^*$ in \cref{fig: optimal sols fixed psi DV and ToF vs phi}.
The values of $\varphi^*$ are here unwrapped to show the unfolding of the connection to the \gls{po} family in a multi-revolution fashion. For the abacus introduced in \cref{sec: abacus of POs}, actual values are always $\varphi \in [0, 2\pi]$.
Instead, \cref{fig: optimal sols fixed psi DVs vs ToF} illustrates the total transfer cost $\Delta v^*$ as well as the individual maneuver components $\Delta v_0^*$ and $\Delta v_f^*$ as functions of the time of flight $\text{ToF}^*$. One notable feature in this figure is that $\Delta v_0^*$ is consistently smaller than $\Delta v_f^*$, and it often approaches zero. This behavior reflects a structural limitation of the current seeding strategy, where the initial guess implicitly assumes $\Delta v_0 = 0$, which in turn biases the solver toward solutions where the initial maneuver is minimal.
This seeding dependence is compounded by the strong nonlinearities of the cislunar dynamical environment, which can occasionally halt the optimization process in suboptimal regions of the solution space, particularly those clustered around $\Delta v_0^* \to 0$.
Despite these limitations, the method remains robust in practice, consistently generating a rich and diverse set of locally optimal solutions across the entire capture set. In the vast majority of cases, the optimizer successfully converges to a local minimum, highlighting the method's effectiveness as a transfer design tool even in the presence of strong dynamical nonlinearities.
To mitigate this seeding bias, a complementary strategy could be employed in which the roles of the maneuvers are reversed, by initializing the optimization with $\Delta v_0 = \Delta v_{\text{mono}}$ and $\Delta v_f = 0$. 
This alternative approach would balance the current preference for minimal initial corrections and could recover many additional solutions. However, adopting such a strategy would require tailored algorithmic adaptations, which are beyond the scope of this work. Additionally, it would roughly double the overall computational cost, while the resulting solutions are expected to follow similar cost and transfer time trends, offering limited practical benefit in most cases.

\begin{algorithm}[tb]
\caption{Spanning the arrival phase $\varphi$ on the \gls{po}.}
\label{alg: algorithm span phase PO}
\begin{algorithmic}[1]
\State{For a given node $i$ on the \gls{bc}, set $\mathbf{x}_0=\mathbf{x}_{BC}(\psi_i)$, as in \cref{eq: initial and final states}. Set $\text{iter}=0$ and $\text{iter}_{max}=20$.} 
\For{$j \gets i+1$ to $n-1$}
    \State{Initialize $p_f=p_j$, $\varphi_f=\varphi_j$, and $\text{ToF}=t(\psi_j)-t(\psi_i)$ using the seeding proposed in \cref{sec: mono-imp cost to family}.}
    \State{Execute lines~\ref{linealg: start while alg core}-\ref{linealg: end while alg core} in \cref{alg: algorithm follow minimum} and store the optimal solution $\Delta v^*$, $p^*$, $\varphi^*$, $\text{ToF}^*$, $\Delta\mathbf{v}_0^*$, and $\Delta\mathbf{v}_f^*$.}
\EndFor
\State{A set of solutions describing the (local optimal) bi-impulsive transfer for each of the $j$-th nodes is stored.}
\State{Delete possible duplicated solutions (the \gls{po} phase ranges with $\varphi \in [\varphi_{j-1}, \varphi_{j+1}]$ overlap).}
\end{algorithmic}
\end{algorithm}

\section{Results for planar transfers} \label{sec: planar results sec}
This section presents the results of the proposed optimization process for in-plane bi-impulsive transfers. The analysis begins with transfer solutions connecting a sample \gls{bc}~$\#1$ at $\Gamma = 0.84$ to the \gls{dro} family, providing a first demonstration of the optimization framework. The study is then extended to include all \glspl{bc} within the same capture set $\mathcal{C}(\Gamma = 0.84)$, and subsequently to investigate how transfer characteristics evolve as $\Gamma$ varies across different capture sets $\mathcal{C}(\Gamma)$. A dedicated Pareto front analysis is employed to compare the resulting transfers from multiple \glspl{bc}. Finally, transfers originating from identical departure nodes on the same \glspl{bc} and inserting into the Lyapunov $L1$ and $L2$ families are addressed, thereby demonstrating the proposed approach's ability to establish consistent and low-cost connections between \glspl{bc} and various \gls{po} families.

\subsection{Sample results for all the departure nodes on a BC} \label{sec: alg entire BC results}

For each departure phase $\psi$ on the current \gls{bc}, we seed the bi‐impulsive solver and span through all the subsequently arrival nodes, as introduced in \cref{sec: BC introduction} and \cref{fig: sample 1 nodes}.  Algorithm \ref{alg: for every BC IC} summarizes this per‐node sweep, whose output is a set of locally optimal solutions.

\begin{algorithm}[tb]
\caption{Analyzing all the departure phases $\psi$ on a \gls{bc}.}
\label{alg: for every BC IC}
\begin{algorithmic}[1]
\For{$i \gets 1$ to $n_0$}
    \State{For the current node $i$ on the \gls{bc}, set $\mathbf{x}_0=\mathbf{x}_{BC}(\psi_i)$, as in \cref{eq: initial and final states}.} 
    \State{Execute \cref{alg: algorithm span phase PO} and store each solution returned by \cref{alg: algorithm follow minimum}.} 
\EndFor
\end{algorithmic}
\end{algorithm}

\begin{figure}[tb]
    \begin{subfigure}[t!]{0.46\textwidth}
        \centering
        \includegraphics[width=\textwidth]{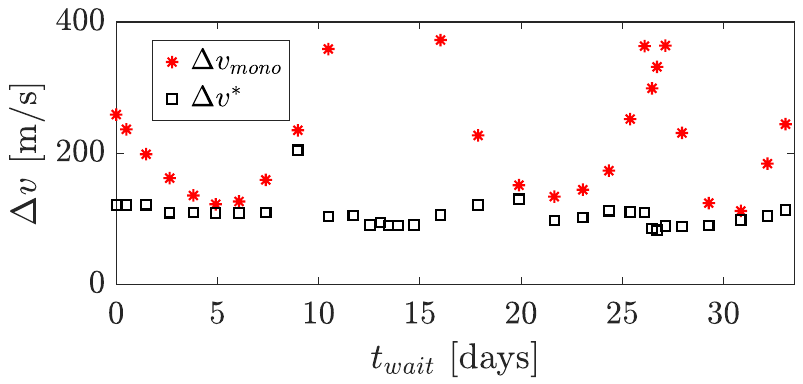}
        \subcaption{Minimum cost $\Delta v^*$ from each of the $n_0$ \gls{bc} departure nodes}
        \label{fig: overall BC sols Dv vs twait}
    \end{subfigure} \hfill
    \begin{subfigure}[t!]{0.46\textwidth}
        \includegraphics[width=\textwidth]{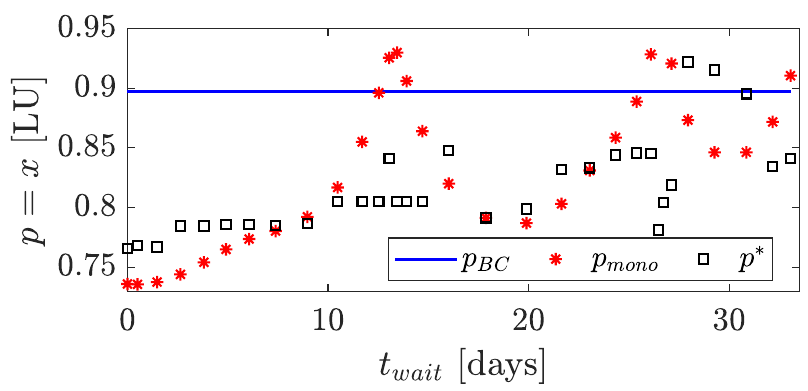}
        \subcaption{Optimal arrival parameter $p^*$ for each of the $n_0$ \gls{bc} departure nodes}
        \label{fig: overall BC sols p vs twait}
    \end{subfigure}
    \\
    \begin{subfigure}[t!]{0.50\textwidth}
        \centering
        \includegraphics[width=\textwidth]{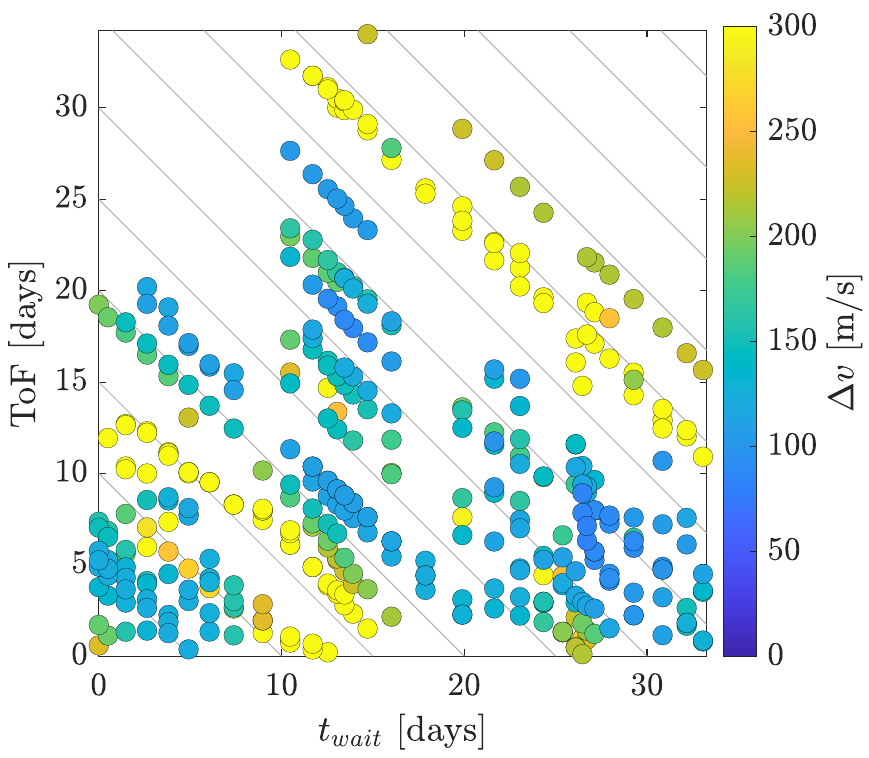}
        \subcaption{Minimum cost $\Delta v^*$ and $\text{ToF}^*$ for each transfer}
        \label{fig: overall BC sols Dv vs ToF vs twait}
    \end{subfigure} \hfill
    \begin{subfigure}[t!]{0.47\textwidth}
        \includegraphics[width=\textwidth]{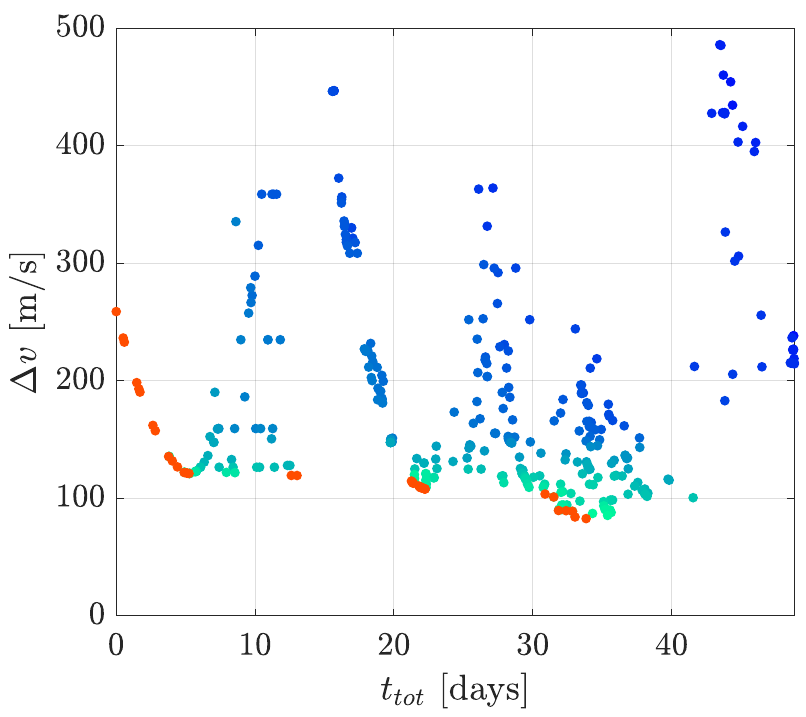}
        \subcaption{Pareto front of cost $\Delta v^*$ vs total time $t_{tot}=t_{wait}+\text{ToF}^*$ for each transfer}
        \label{fig: overall BC sols pareto}
    \end{subfigure}
    \caption{Overall results for sample \gls{bc}~$\#1$ generated using \cref{alg: for every BC IC}.}
    \label{fig: overall BC sols}
\end{figure}

The results obtained from \cref{alg: for every BC IC} represent the complete set of (local optimal) bi-impulsive transfers from each of the $i$-th departure nodes to each of the $j$-th arrival nodes. All transfers from a given \gls{bc} to a family of \glspl{po} are stored in a structured set.
Cases where $i = j$ correspond to the mono-impulsive solutions introduced in \cref{sec: mono-imp cost to family}. Some combinations of $i$ and $j$ may be missing due to non-convergence or overlapping solutions in the bi-impulsive method.
Results for sample \gls{bc}~$\#1$ are shown in \cref{fig: overall BC sols}, where each of the $n_0$ \gls{bc} departure nodes corresponds to an implicit waiting time $t_{wait}(\psi_i)$, with $i = 1, 2, \dots, n_0$. As such, the phase $\psi_i$ serves both to identify the departure node and to indicate the timing of the associated transfer.

Figs.~\ref{fig: overall BC sols Dv vs twait} and \ref{fig: overall BC sols p vs twait} show only the minimum cost solution from each departure node $\mathbf{x}_i$. They represent, respectively, the minimum overall cost $\Delta v^*$ for a certain $\mathbf{x}_i$ and the corresponding parameter on the arrival \gls{dro} family $p^*$ as a function of the waiting time $t_{wait}$.
These figures also represent the mono-impulsive cost $\Delta v_{mono}$ and parameters $p_{mono}$ as obtained in \cref{sec: mono-imp cost to family}.
The parameter $p$ used to parametrize this \gls{dro} family is $p=x$, where $x$ is the coordinate at the Poincaré section in $y=0$ and $\dot{y}>0$. The value $p_{BC}$ shown in \cref{fig: overall BC sols p vs twait} is obtained by retrieving the Jacobi constant $C_J$ of the current \gls{bc} (obtained inverting \cref{eq: gamma} using $\Gamma=0.84$) and then computing the value of $x$ in the Poincaré section which provides this value of $C_J$.
Note that \cref{fig: overall BC sols p vs twait} shows the tendency of arrival \glspl{po} to have $p^*<p_{BC}$, which means that higher three-body energies \glspl{dro} are more likely to be reached. Instead, with increasing waiting time $t_{wait}$, the mono-impulsive $p_{mono}$ tends to oscillate more closely around the value $p_{BC}$. In the same fashion, the value for $p^*$ tends to increase, leading to arrival \glspl{dro} that are more heavily bound to the Moon. In fact, for the value $p=p_{BC}$, the \gls{dro} are contained well within the Hill's sphere of the Moon, as displayed in \cref{fig: best transfer fixed psi traj}.
Note that the parameter $p$ can always be translated in terms of three-body energy $\Gamma$ or Jacobi constant, leading to $\Delta C_J = C_{J, PO} - C_{J, BC}$.
These considerations suggest that, to favor final \glspl{po} more tightly bound to the Moon, a constraint on the Jacobi constant (i.e., the family parameter $p$) could be included to force an increased $C_J$ value for the solutions.

All the solutions of \cref{alg: for every BC IC} are provided in \cref{fig: overall BC sols Dv vs ToF vs twait}, where colored markers represent the optimal cost $\Delta v^*$ in a $\text{ToF}^*$ against $t_{wait}$ graph.
Here, patterns highlighted by the gray diagonal lines with equation $t_{wait}+\text{ToF}=t_{tot}=\text{const}$ are clearly visible.
In addition, areas with clustered solutions can be identified, separated by regions where the algorithm's convergence tends to fail. For instance, this occurs at $t_{wait}+\text{ToF}\approx 15$ days and $t_{wait}+\text{ToF}\approx 27$ days, where the mono-impulsive seed solutions $\Delta v_{mono}$ in \cref{fig: overall BC sols Dv vs twait} tend to diverge. This is particularly true for $t_{wait}+\text{ToF}\approx 15$ days, when sample \gls{bc}~$\#1$  is in a prograde phase of the trajectory resembling a Lyapunov $L1$ \gls{po} (see \cref{fig: sample 1 trajectory} and \cref{sec: BC introduction}). Finally, another region with high costs and non-convergence of the solution is evident for $t_{wait}+\text{ToF}\approx 45$ days, when the capture phase of sample \gls{bc}~$\#1$ has almost come to an end and the trajectory is close to escaping.

All the solutions of \cref{alg: for every BC IC} are also represented in \cref{fig: overall BC sols pareto} in terms of cost $\Delta v^*$ against total transfer time $t_{tot}=t_{wait}+\text{ToF}^*$, with Pareto front solutions highlighted in red.
The total computational cost of the C++ optimization process for transfers from a given \gls{bc} to a specific \gls{po} family is typically on the order of 10 minutes when executed on a single-core, standard desktop machine.

\subsection{Pareto front analysis across BCs and energy levels} 
\label{sec: pareto comparison between BCs to DRO}
\begin{figure}[tbp]
    \centering
    \begin{subfigure}[t!]{0.49\textwidth}
        \centering
        \includegraphics[width=\textwidth]{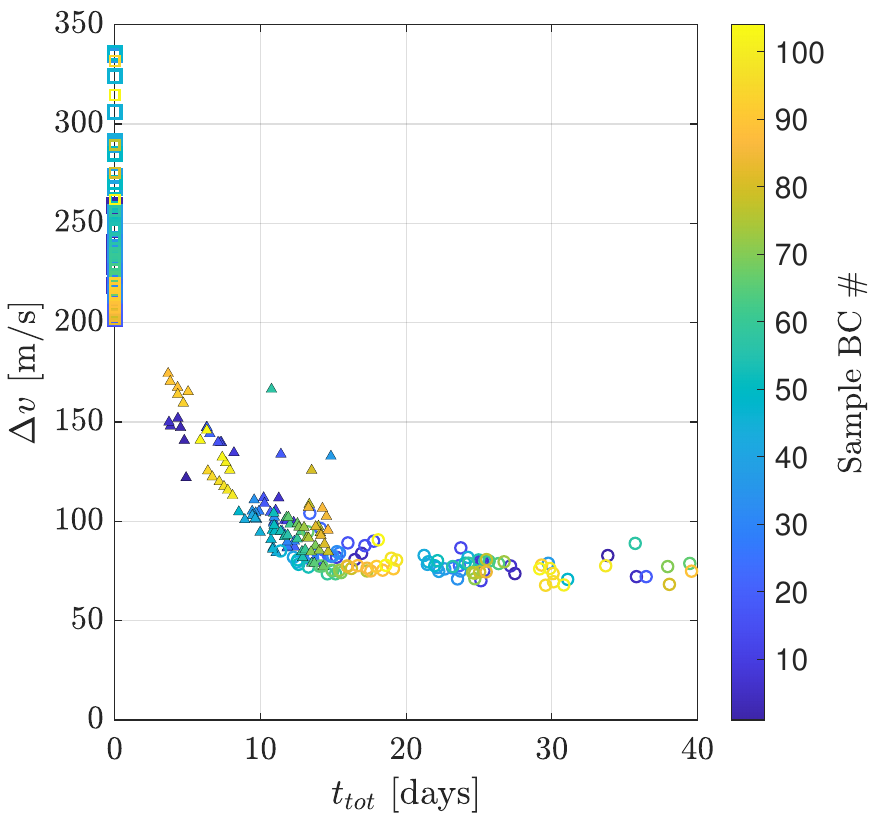}
        \subcaption{Main features of the Pareto front for  $\Gamma=0.84$}
        \label{fig: pareto front features gamma84}
    \end{subfigure} \hfill
    \begin{subfigure}[t!]{0.49\textwidth}
        \includegraphics[width=\textwidth]{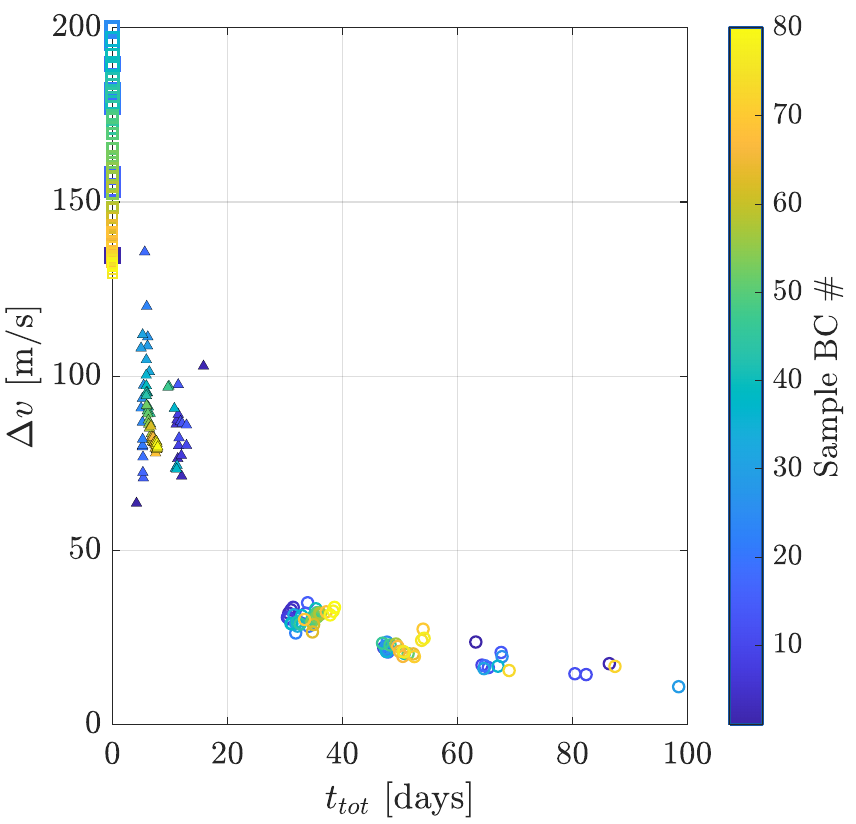}
        \subcaption{Main features of the Pareto front for  $\Gamma=1.18$}
        \label{fig: pareto front features gamma118}
    \end{subfigure}
    \\
    \begin{subfigure}[t!]{0.49\textwidth}
        \centering
        \includegraphics[width=\textwidth]{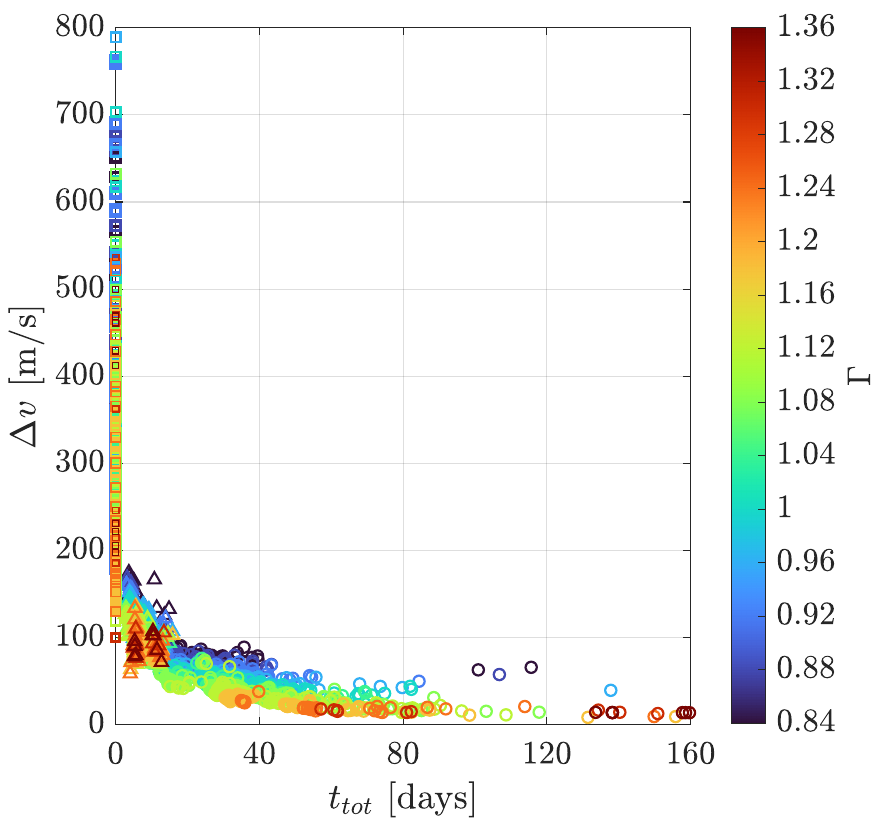}
        \subcaption{Main features of the Pareto front for varying $\Gamma$}
        \label{fig: pareto front features gamma variable}
    \end{subfigure} \hfill
    \begin{subfigure}[t!]{0.49\textwidth}
        \includegraphics[width=\textwidth]{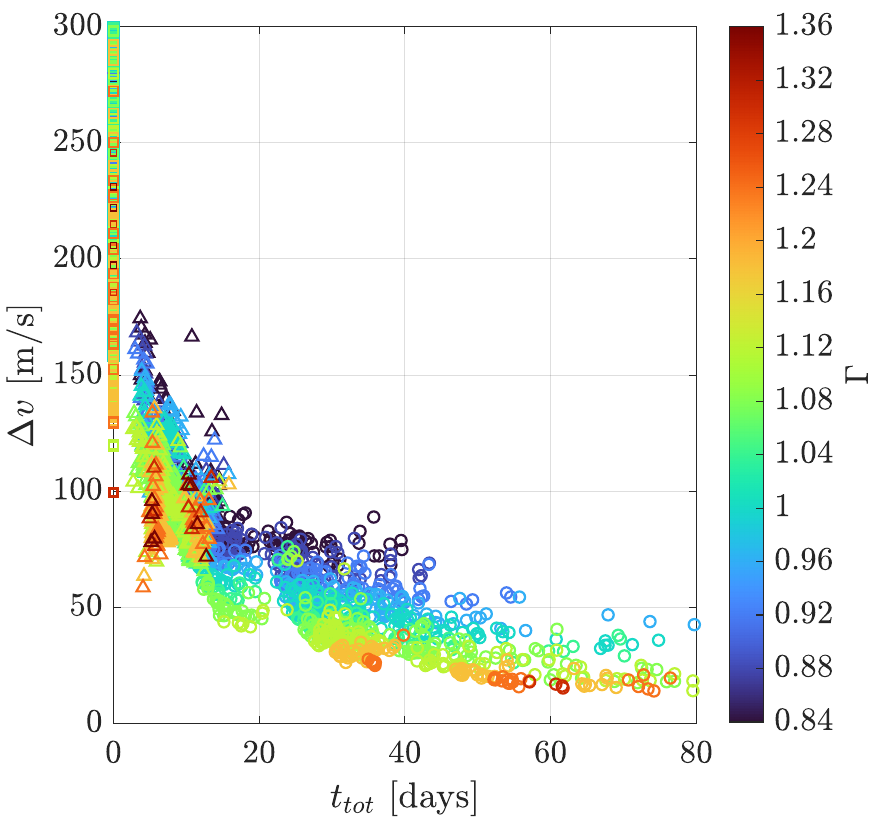}
        \subcaption{Close-up of \cref{fig: pareto front features gamma variable}}
        \label{fig: pareto front features gamma variable close up}
    \end{subfigure}
    \caption{Main features ($t_{tot}=0$, \textit{knee}, and maximum $t_{tot}$) of the Pareto front.}
    \label{fig: pareto front features}
\end{figure}

The optimization procedure described in \cref{sec: alg entire BC results} is extended to all the sample \glspl{bc} of $\mathcal{C}(\Gamma = 0.84)$ highlighted with yellow circles in \cref{fig: sample BCs on capture set}. The resulting Pareto fronts are summarized in \cref{fig: pareto front features gamma84}, where three representative points are extracted from each complete. These points correspond to the mono-impulsive solution at $t_{tot} = 0$ (square marker), the (lowest-cost) solution at the maximum transfer time $t_{tot}$ (circle marker), and the \textit{knee} of the Pareto front (triangle marker), defined as the point closest to the origin in the $(\Delta v/10, \, t_{tot})$ plane.
\cref{fig: pareto front features gamma118} displays the same three Pareto front features, this time computed for 80 representative \glspl{bc} sampled from the set $\mathcal{C}(\Gamma = 1.18)$. These 80 \glspl{bc} again constitute the $0.01\%$ of the 2+ retrograde revolutions subset of the capture set at this higher three-body energy level. Note the different axes scale in this second figure, which highlights the availability of much lower costs and much longer transfer times.
This feature is analyzed in detail in \cref{fig: pareto front features gamma variable}, where a comparison of the Pareto front features across different energy levels is provided. A detailed view of the same figure is shown in \cref{fig: pareto front features gamma variable close up}. These plots clearly demonstrate the influence of the three-body energy parameter $\Gamma$ on the transfer performance. As expected, higher values of $\Gamma$ tend to correspond to lower-cost insertions into the \gls{dro} family. This trend is consistent with the structure of the stability regions introduced in \cref{sec: BCs and POs poincare} and with the mono-impulsive cost estimates discussed in \cref{sec: mono-imp cost estimate poincare}. At $\Gamma = 1.18$, the \gls{dro} stability region is narrowly concentrated around the central \gls{dro}, favoring cheaper insertions. In contrast, for $\Gamma = 0.84$, the \gls{dro} stability region extends more broadly in the $x$–$\dot{x}$ phase space, requiring more expensive insertion maneuvers.
Finally, the less predictable behavior and higher $t_{tot}$ values observed at higher energy levels (toward the red) stem from the interplay between the \gls{bc} search method of~\cite{BC-ETDjournal} and the energy properties of the resulting trajectories. As $\Gamma$ increases, both the three-body and two-body energy levels grow, affecting the structure and longevity of the identified \glspl{bc}. For further details, see Sections V.C and VI.A.3 of~\cite{BC-ETDjournal}.

\subsection{Transfers to Lyapunov families} \label{sec: results to Lyap} 
We now apply the transfer optimization method to the Lyapunov $L1$ and Lyapunov $L2$ families, mirroring the approach adopted for the \gls{dro} family. In this case, departure nodes on the \gls{bc} are restricted to the semi-region closest to the respective libration point. Specifically, for Lyapunov $L1$ transfers, the previously introduced nodes are considered only until the first instance where $x_i > 1 - \mu$; for Lyapunov $L2$, the process stops at the first $x_i < 1 + \mu$. The total number of nodes $n_0 = n$ is thus determined by this truncation of the full \gls{bc}.
As a result, only a subset of the sample \glspl{bc} considered in the previous section is suitable for targeting a given Lyapunov family. Specifically, the transfer method is applied to the Lyapunov $L1$ family only if the \gls{bc} approaches from the $L1$ side of the synodic position space, and analogously for Lyapunov $L2$. While opportunities for insertion into a Lyapunov orbit may arise later along the \gls{bc}, such transfers are excluded from the present analysis due to the high $t_{wait}$ values (and therefore longer total transfer durations $t_{tot}$) they would entail.

\begin{figure}[tb]
    \begin{subfigure}[t!]{0.46\textwidth}
        \centering
        \includegraphics[width=\textwidth]{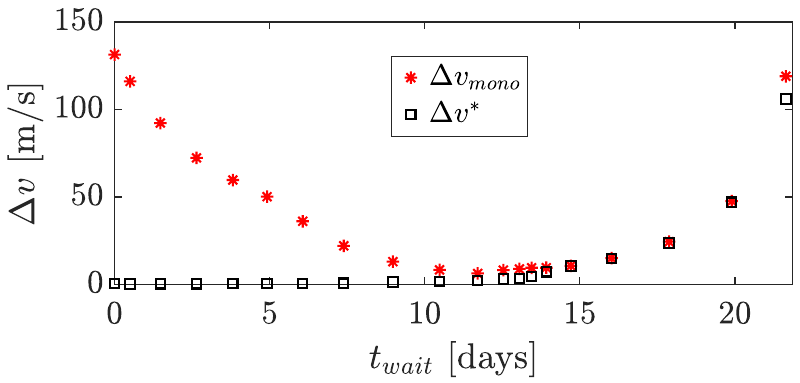}
        \subcaption{Minimum cost $\Delta v^*$ from each of the $n_0$ \gls{bc} departure nodes}
        \label{fig: overall BC sols Dv vs twait L1}
    \end{subfigure} \hfill
    \begin{subfigure}[t!]{0.46\textwidth}
        \includegraphics[width=\textwidth]{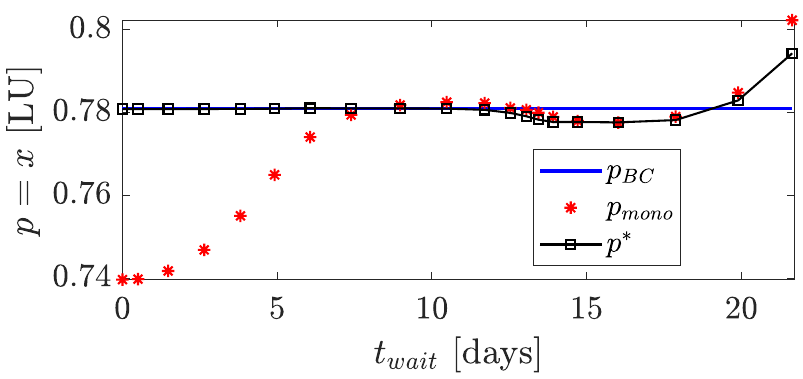}
        \subcaption{Optimal arrival parameter $p^*$ for each of the $n_0$ \gls{bc} departure nodes}
        \label{fig: overall BC sols p vs twait L1}
    \end{subfigure}
    \\
    \begin{subfigure}[t!]{0.49\textwidth}
        \centering
        \includegraphics[width=\textwidth]{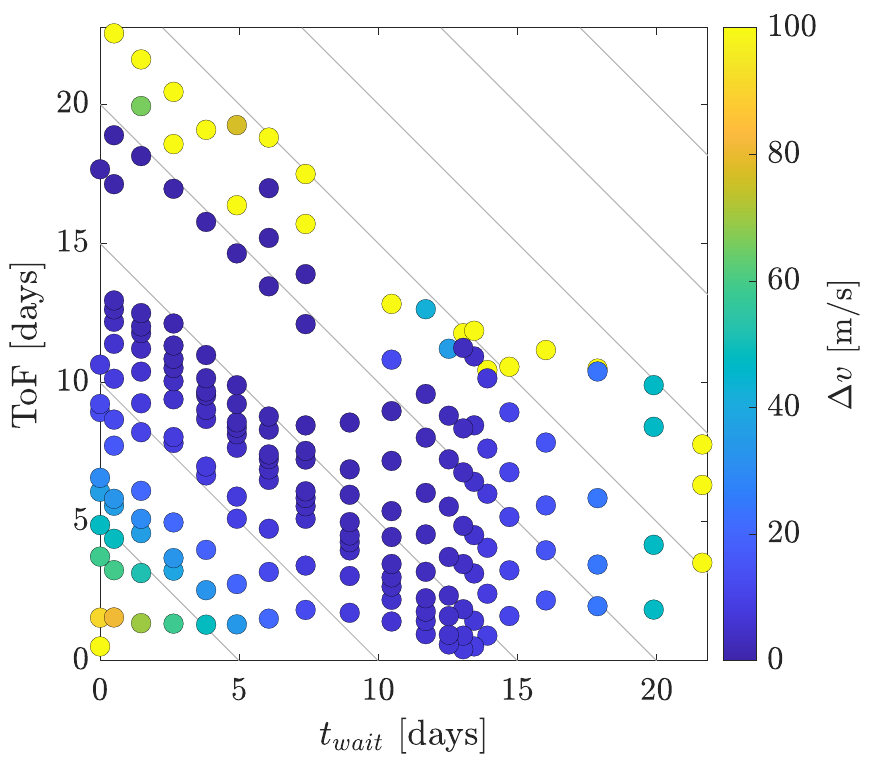}
        \subcaption{Lowest cost $\Delta v^*$ and $\text{ToF}^*$ for each transfer}
        \label{fig: overall BC sols Dv vs ToF vs twait L1}
    \end{subfigure} \hfill
    \begin{subfigure}[t!]{0.48\textwidth}
        \includegraphics[width=\textwidth]{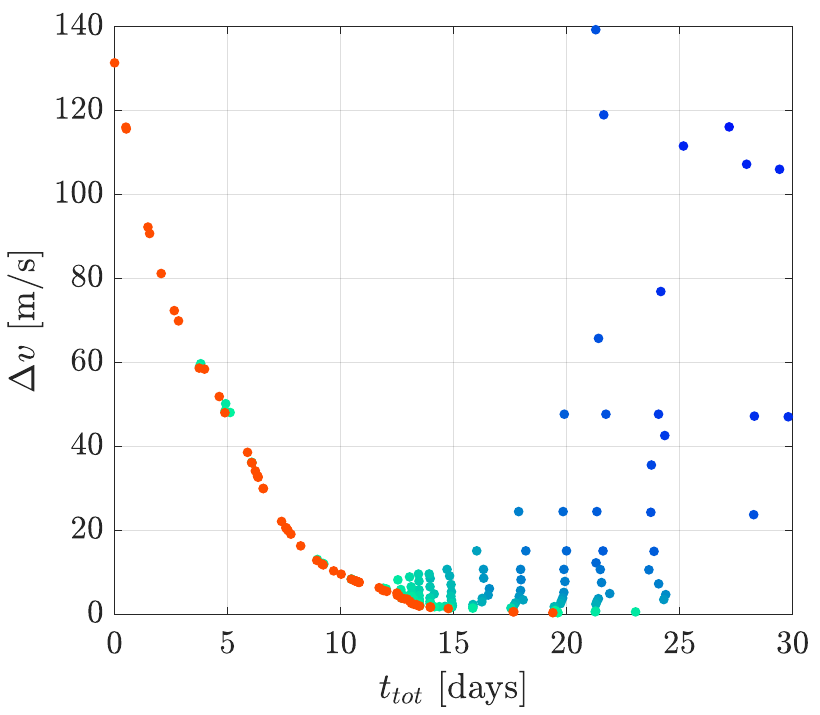}
        \subcaption{Pareto front of cost $\Delta v^*$ vs total time $t_{tot}=t_{wait}+\text{ToF}^*$ for each transfer}
        \label{fig: overall BC sols pareto L1}
    \end{subfigure}
    \caption{Transfer results from sample \gls{bc}~$\#1$ for insertion into the Lyapunov $L1$ family.} 
    \label{fig: overall BC sols L1}
\end{figure}

\begin{figure}[tb]
    \begin{subfigure}[t!]{0.49\textwidth}
        \centering
        \includegraphics[width=\textwidth]{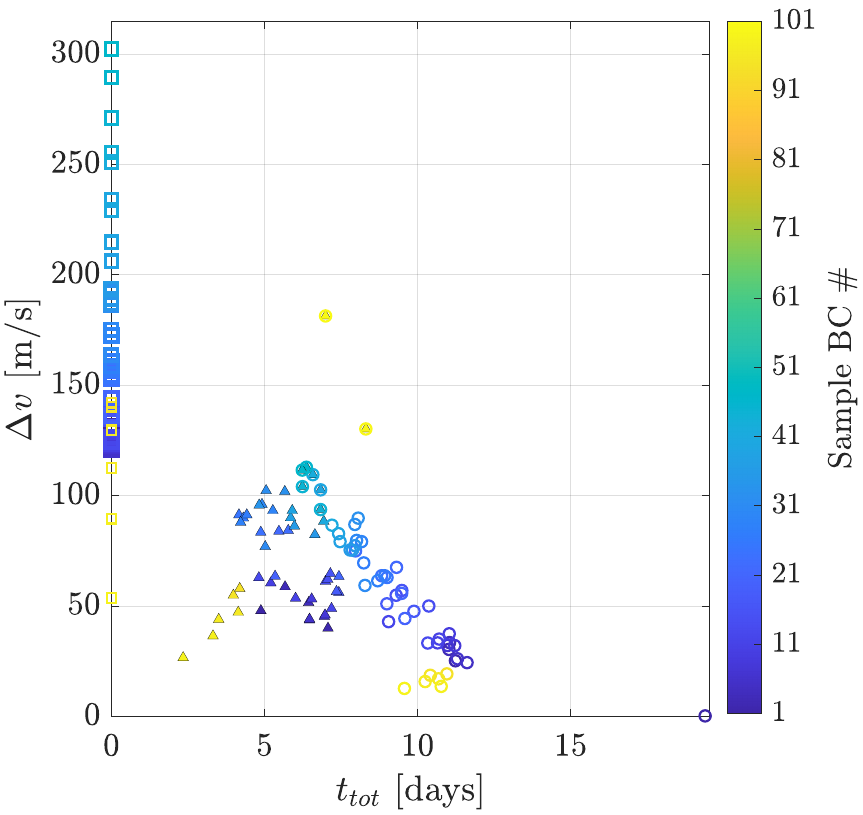}
        \subcaption{Main features of the Pareto front for $\Gamma=0.84$}
        \label{fig: pareto front features gamma84 L1}
    \end{subfigure} \hfill
    \begin{subfigure}[t!]{0.49\textwidth}
        \centering
        \includegraphics[width=\textwidth]{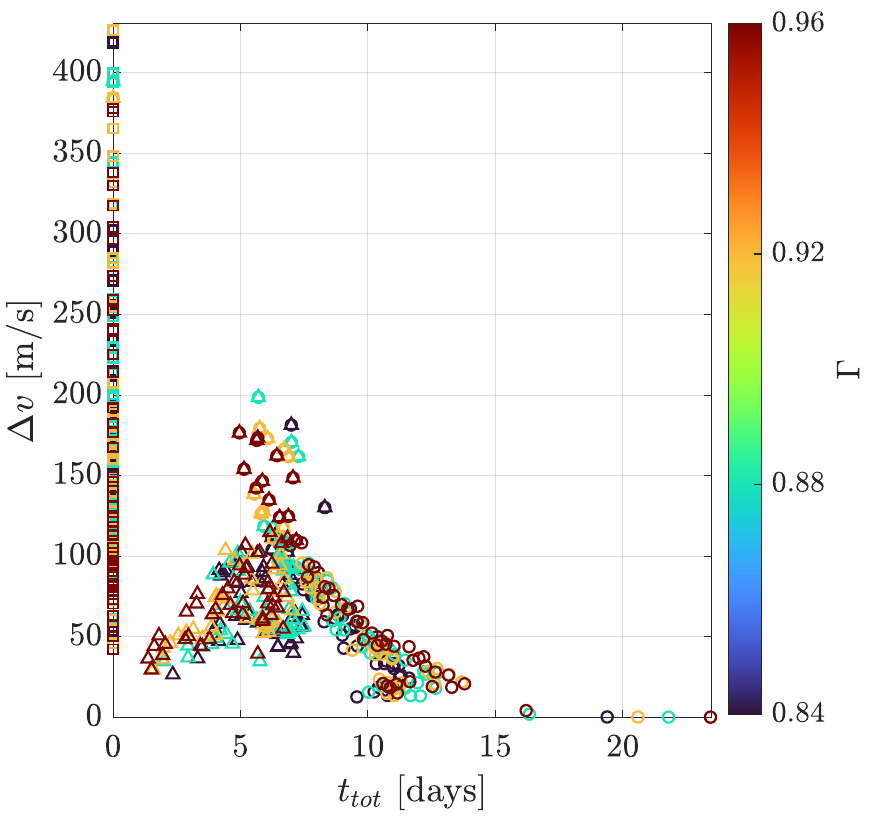}
        \subcaption{Main features of the Pareto front for varying $\Gamma$}
        \label{fig: pareto front features gamma variable L1}
    \end{subfigure}
    \caption{Main features ($t_{tot}=0$, \textit{knee}, and maximum $t_{tot}$) of the Pareto front analysis for insertion into the Lyapunov $L1$ family.}
    \label{fig: pareto front features L1}
\end{figure}

The resulting transfer characteristics for the Lyapunov $L1$ family are summarized in Figs.~\ref{fig: overall BC sols L1} and \ref{fig: pareto front features L1}, which mirror the structure of the results presented earlier for the \gls{dro} family.
\Cref{fig: overall BC sols Dv vs twait L1} reveals that the lowest-cost transfers are achieved at early departure times, specifically for $t_{wait} < 12$ days. In this regime, the \gls{bc} trajectory naturally approaches the Lyapunov $L1$ \gls{po}, enabling efficient insertions. For $t_{wait} > 12$ days, instead, the \gls{bc} evolves toward a \gls{dro}-type dynamics, leading to an increase in the required $\Delta v^*$. This transition is also reflected in \cref{fig: overall BC sols p vs twait L1}, where the arrival parameter $p^*$ remains near $p_{BC}$ for early transfers but begins to diverge as $t_{wait}$ increases. \Cref{fig: overall BC sols Dv vs ToF vs twait L1} reinforces these observations and introduces an implicit constraint on the maximum total transfer time, suggesting that $t_{tot} < 20$ days is a practical upper bound. Indeed, all Pareto-optimal solutions highlighted in red in \cref{fig: overall BC sols pareto L1} fall below this threshold.

\Cref{fig: pareto front features gamma84 L1} displays the three Pareto front features for transfers departing from each \gls{bc} in a subset of $\mathcal{C}(\Gamma = 0.84)$, whose trajectories originate on the $L1$ side.
Note that these \glspl{bc} were selected based on their completion of 2 or more retrograde revolutions, hence a reasonable comparison with the \gls{dro} insertion features is possible. Nonetheless, many additional \glspl{bc} exhibiting a broader range of characteristics could be extracted from $\mathcal{C}(\Gamma = 0.84)$ depending on specific mission objectives.
In contrast, \cref{fig: pareto front features gamma variable L1} presents the same analysis extended across multiple three-body energy levels. Unlike the case of \gls{dro} insertions, these results show that variations in the energy parameter $\Gamma$ have little effect on the insertion cost into the Lyapunov $L1$ family. This insensitivity suggests that the local dynamics near the $L1$ point remain largely unchanged across the energy levels considered, in contrast to the more pronounced dependence observed in the \gls{dro} insertion case. This behavior is to be attributed to the absence of a stability region around the Lyapunov $L1$ orbits, as opposed to the \gls{dro} stability region.

\begin{figure}[tb]
    \centering
        \begin{subfigure}[t!]{0.43\textwidth}
            \raggedright
            \includegraphics[width=\textwidth]{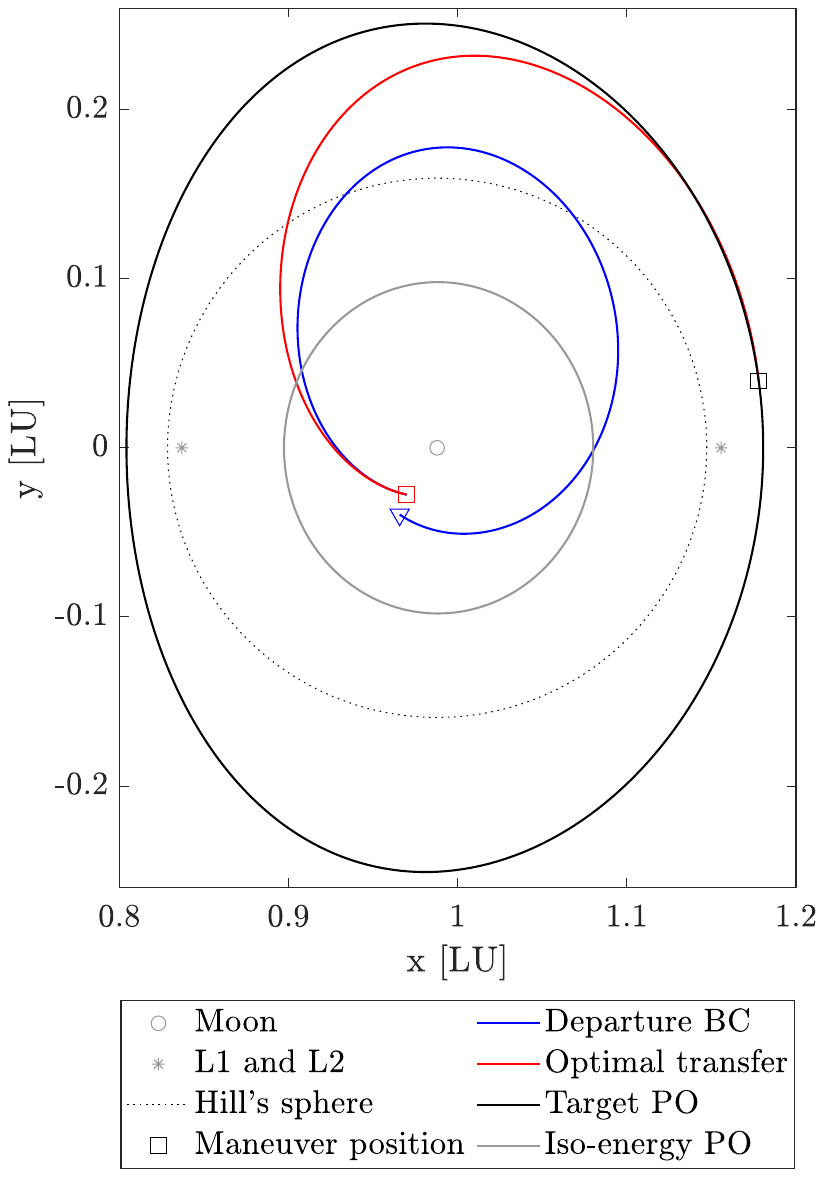}
            \subcaption{Optimized transfer from node $i=26$ to \gls{dro}.}
            \label{fig: best transfer fixed psi traj}
        \end{subfigure}
        \hfill
        \begin{subfigure}[t!]{0.25\textwidth}
            \centering
            \includegraphics[width=0.7\textwidth]{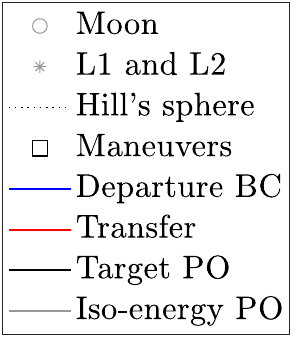}
        \\
        \vspace{1.7cm}
            \includegraphics[width=\textwidth]{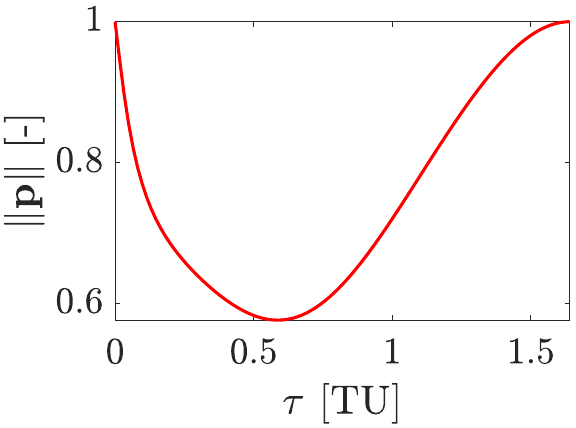}
            \subcaption{Primer vector $\|\mathbf{p\|}$ analysis~\cite{jezewski1975PrimerVector} for transfer in \cref{fig: best transfer fixed psi traj}}
            \label{fig: best transfer fixed psi PV}
        \end{subfigure}
        \hfill
        \begin{subfigure}[t!]{0.245\textwidth}
            \raggedleft
            \includegraphics[width=\textwidth]{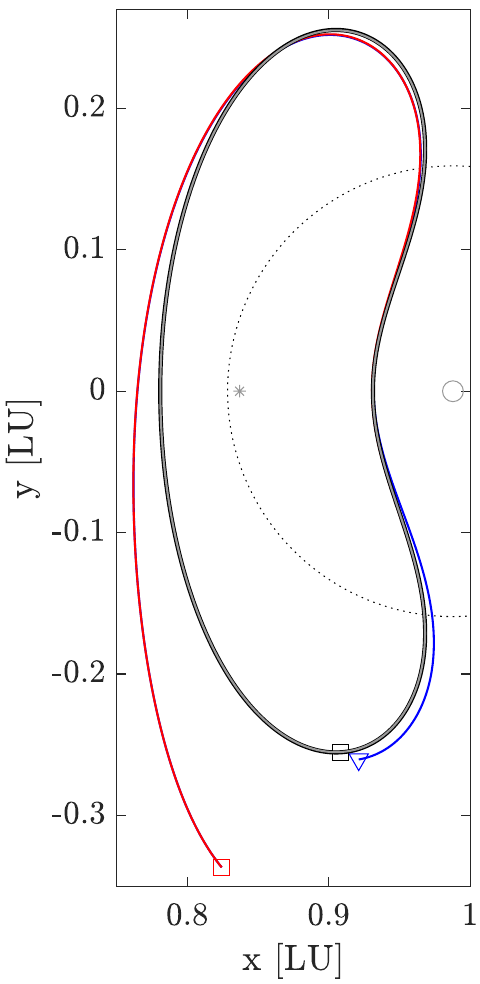}
            \subcaption{Optimized transfer from node $i=2$ to Lyapunov $L1$.}
            \label{fig: best transfer Lyap L1}
        \end{subfigure}
        \caption{Best transfers from sample \gls{bc}~$\#1$ in \cref{fig: sample 1 trajectory} to \gls{dro} and Lyapunov $L1$ families.}
        \label{fig: best transfer fixed psi}
\end{figure}

The bi-impulsive transfer to the \gls{dro} and Lyapunov $L1$ families having minimum overall cost $\Delta v$ are shown in \cref{fig: best transfer fixed psi}.
In particular, \cref{fig: best transfer fixed psi traj} shows a transfer with the lowest cost from \cref{fig: optimal sols fixed psi}. It originates from departure node $i = 26$ and achieves a time of flight $\text{ToF}^* = 7.12$ days with a total cost of $\Delta v^* = 82.77$ m/s. Additionally, \cref{fig: best transfer fixed psi PV} shows the primer vector~\cite{jezewski1975PrimerVector} of this transfer, proving that additional intermediate impulses do not improve this bi-impulsive solution.
The transfer to the Lyapunov $L1$ family, instead, is represented in \cref{fig: best transfer Lyap L1}. It originates from departure node $i = 2$ and achieves a time of flight $\text{ToF}^* = 17.13$ days with a total cost of $\Delta v^* = 0.6$ m/s.

For brevity, optimal transfer solutions for insertion into the Lyapunov $L2$ family are not shown, as they follow trends and exhibit characteristics very similar to those of the Lyapunov $L1$ family. This similarity is consistent with the well-known dynamical symmetry of the two Lyapunov families, the time-reversal symmetry of \gls{cr3bp} trajectories, and the comparable number of \gls{bc} trajectories connecting to (or departing from) each side of the Moon~\cite{BC-ETDjournal}.

\section{Extension to the spatial problem} \label{sec: spatial problem}

This section extends the proposed methodology to the spatial case. The optimization framework developed in this work is implemented in a general form, allowing for a straightforward extension to spatial \glspl{bc} and spatial \gls{po} families with only minor modifications. While the overall optimization process remains applicable, the seeding procedure based on mono-impulsive solutions must be adapted. As introduced in \cref{sec: mono-imp cost to family}, in the planar case, each position along a \gls{bc} can be directly associated with a \gls{dro} or Lyapunov \gls{po}; this correspondence no longer holds in the spatial setting, where planar families lie on a four-dimensional subspace of the six-dimensional \gls{cr3bp} phase space.

In this context, the section first describes the selection process used to identify promising spatial \glspl{bc} from the database developed in~\cite{BC-3Djournal}, focusing on captures exhibiting characteristics compatible with the targeted halo (specifically, \glspl{nrho}) and butterfly families~\cite{caleb2023}. A dedicated procedure for adapting the seeding algorithm to the spatial case is then introduced. The resulting optimized bi-impulsive transfers from the selected \glspl{bc} to the spatial \gls{po} families are presented and analyzed, highlighting the dynamical properties of the selected captures. Representative trajectories are discussed in detail to illustrate the overall performance and the spatial optimization framework's ability to consistently generate reliable connections to the targeted families.

\subsection{Spatial BCs selection procedure} \label{sec: subset spatial Gamma90}

While the planar capture sets $\mathcal{C}(\Gamma)=\mathcal{C}(\Gamma, z=0, \zeta=0)$~\cite{BC-ETDjournal} contain a limited number $n_C$ of \glspl{bc}, the spatial sets $\mathcal{C}(\Gamma, z, \zeta)$~\cite{BC-3Djournal} include a significantly larger population, typically on the order of $(n_C)^2$ for the same value of $\Gamma$. This increase in dimensionality makes the selection of specific subsets within $\mathcal{C}(\Gamma, z, \zeta)$ particularly critical.
As we want to focus on \glspl{nrho} and butterfly \glspl{po}, the goal is to isolate trajectories that exhibit characteristics favorable for insertion into these families. To this end, the selection constraints are tailored to reproduce the geometric and dynamical features of the target \gls{po} families—such as their inclination and periapsis location around the Moon. A similar approach was proposed in~\cite{BC-3Djournal}, where analogous geometric conditions were applied through the following set of constraints:
\begin{itemize}
    \item \glspl{bc} must complete at least two revolutions around the Moon, as required by the planar criteria in \cref{sec: BC introduction};
    \item Minimum perilune distance: $r_{2,\min} < 10\,R_M$, where $R_M=1737.4$ km is the Moon’s physical radius;
    \item Inclination at perilune: $\left| i_{2,\min} - 90^\circ \right| < 6^\circ$;
    \item Argument of perilune: $\left| \omega_{2,\min} - 90^\circ \right| < 12^\circ$.
\end{itemize}
For example, with these restrictions, the approximately $10^8$ \glspl{bc} in $\mathcal{C}(\Gamma=0.90, z, \zeta)$ are effectively reduced to about 200 trajectories exhibiting geometrical features of \gls{nrho} and butterfly orbits. For computational reasons, 50 \glspl{bc} are extracted, which uniformly represent the filtered subset~\cite{BC-3Djournal}.

\subsection{Adjustment of the seeding procedure} \label{sec: spatial seeding adjustment}
In the spatial case, the simplified analysis of \cref{sec: mono-imp cost to family} does not hold anymore. Here, we address the spatial adjustment of the seeding procedure assuming that the \gls{bc} dynamics is still governed by one or more underlying \gls{po} families. Accordingly, each \gls{bc} that follows the dynamics of a selected \gls{po} family must intersect the subspace in which that family resides — a hypothesis that is verified a posteriori.

To enable the use of the same optimization framework described in the planar case, spatial intersections in position space must be identified to generate suitable initial seeds. This is accomplished by computing, for each \gls{bc} node $k$ at phase $\psi_k$ returned by the numerical integration of \cref{eq: equations of motion}, the minimum spatial distance $d_k$ to the \gls{po} family:
\begin{equation}
    d_k = \| \mathbf{r}_{PO}(p,\varphi) - \mathbf{r}_{BC,k} \| \, .
    \label{eq: dist BC PO family spatial}
\end{equation}
The search for the closest point and the corresponding \gls{po} family parameters $p$ and $\varphi$ is performed using the same iterative, adaptive-grid method introduced in \cref{sec: mono-imp cost to family}. However, unlike in the planar case, the distance $d_k$ will not generally reach zero for any pair $(p_k, \varphi_k)$. Instead, the phases $\psi_k$ where $d_k$ reaches a local minimum are identified, producing a subset $\psi_q$ of promising intersection phases, where $q \ll k$.

These candidate phases $\psi_q$ are further refined using a \gls{da}-based polynomial expansion of the \gls{bc} dynamics via \cref{eq: dynamics with variable time}, allowing more precise determination of the intersection points in the position space.
Similarly to what introduced in \cref{sec: mono-imp cost to family}, a pair $(p_f, \varphi_f)$ is found (using map inversion) such that $\sqrt{(x_{PO} - x)^2 + (y_{PO} - y)^2 + (z_{PO} - z)^2} = 0$. As a consequence, the entire state $\mathbf{x}_{PO}(p_f, \varphi_f)$ is retrieved, and a mono-impulsive correction for a transfer from $\mathbf{x}_{BC,q}$ to $\mathbf{x}_{PO}$ is computed as
\begin{equation}
    \Delta v_{\text{mono}}=\sqrt{(\dot{x}_{PO}-\dot{x}_{BC,q})^2+(\dot{y}_{PO}-\dot{y}_{BC,q})^2+(\dot{z}_{PO}-\dot{z}_{BC,q})^2} \, .
    \label{eq: deltav mono spatial}
\end{equation}

The $n_0$ departure nodes are selected using the same method as described at the beginning of \cref{sec: opt section} 
In contrast, the $n_q$ refined nodes at phases $\psi_q$ serve as the arrival nodes for the optimization method described in Sections~\ref{sec: core steps}, \ref{sec: follow minimum}, and~\ref{sec: span on phase}. In the spatial case, the number of available arrival nodes is typically much smaller, since intersections between the \gls{bc} and the \gls{po} family subspace occur regularly, though in a limited number, typically on the order of $n_q=n_0/5$. Nonetheless, they provide a sufficient set of arrival nodes to enable multiple distinct transfer solutions. 

Although fewer initial guesses are required, the total computational cost of the C++ optimization process for spatial transfers from a given \gls{bc} to a specific \gls{po} family is considerably higher than in the planar case, typically around 30 minutes on a single-core, standard desktop machine. This increase is primarily due to the additional degrees of freedom involved in the spatial propagation and subsequent optimization steps.


\subsection{Spatial results from BCs to halo and butterfly families} \label{sec: spatial capture sets to all families}

\begin{figure}[tb]
    \raggedright
    \begin{minipage}{.47\textwidth}
        \includegraphics[width=\textwidth]{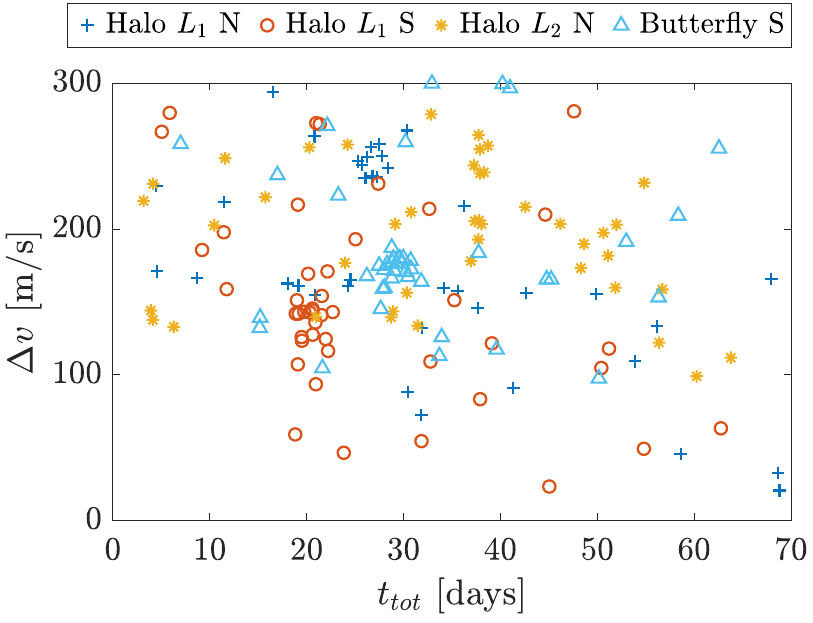}
        \caption{Best transfers from \glspl{bc} of \cref{sec: subset spatial Gamma90} to all families.}
        \label{fig: subset Gamma90 to all fams}
    \end{minipage}
    \hfill
    \raggedleft
    \begin{minipage}{.52\textwidth}
        \includegraphics[width=\textwidth]{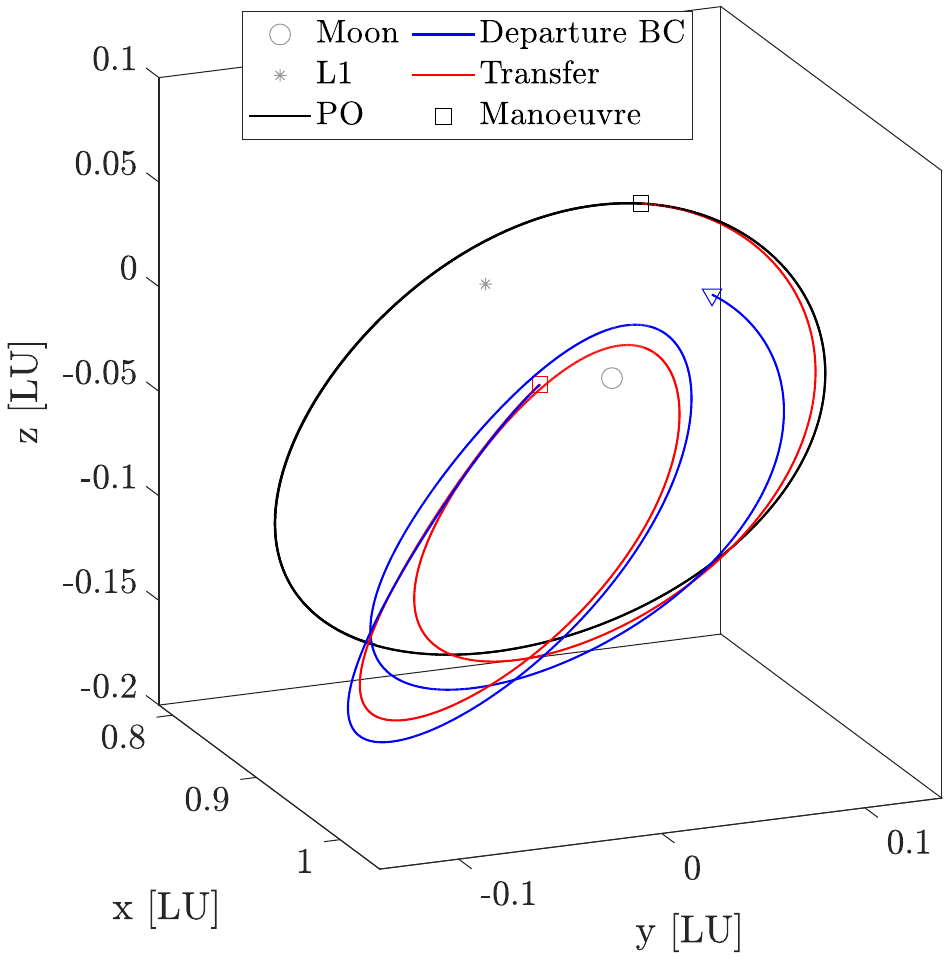}
        \caption{Best transfer from \glspl{bc} of \cref{sec: subset spatial Gamma90} to southern halo $L1$: $\Delta v^*\approx23$ m/s.}
        \label{fig: best sample subset to halo L1 S}
    \end{minipage}
\end{figure}

Results based on the subset of spatial departure \glspl{bc} introduced in \cref{sec: subset spatial Gamma90} are now presented. The optimization process is employed to compute transfers from each of the 50 selected \glspl{bc} to all available families.
In \cref{fig: subset Gamma90 to all fams}, only the most cost-effective arrival families are shown. Transfers toward the southern halo $L2$ and the northern butterfly families consistently result in higher costs for equal transfer times $t_{tot}$.
Overall, the halo $L1$ family appears to offer the most cost-effective insertion options for this subset, with the southern branch standing out in particular due to its lower $t_{tot}$. Conversely, for very short durations ($t_{tot} = 0 \div 18$ days), only a few transfers with reasonable cost are found. As detailed below, this behavior can be attributed to the dominance of Lyapunov-like motion during this early phase of capture.
Interestingly, clusters of solutions targeting the same family emerge at specific times: for example, at $t_{tot} \approx 21$, $27$, $29$, and $37$ days, corresponding to the southern and northern halo $L1$, southern butterfly, and northern halo $L2$ families, respectively. These clusters suggest the presence of \gls{bc} corridors with similar dynamical characteristics.
Finally, as expected, halo families generally provide more favorable insertion opportunities compared to their butterfly counterparts, which is a direct consequence of their more stable dynamical behavior.

The best solution from \cref{fig: subset Gamma90 to all fams} targeting a southern halo $L1$ family is illustrated in \cref{fig: best sample subset to halo L1 S} and features a total transfer time of approximately $t_{tot} \approx 45$ days and a cost of $\Delta v^* \approx 23$~m/s. 
These results demonstrate the method’s ability to efficiently identify low-cost transfer opportunities across a broad range of conditions. In addition, since the selected \glspl{bc} are generated with $z > 0$ only, the symmetry of the \gls{cr3bp} with respect to the $x$–$y$ plane is exploited to effectively double the number of initial \glspl{bc} and transfer options without additional computation. For example, any transfer shown in \cref{fig: subset Gamma90 to all fams} targeting the southern halo $L1$ also implies the existence of a symmetric transfer to the northern halo $L1$. 

\begin{figure}[tb]
    \begin{subfigure}[t!]{0.48\textwidth}
        \centering
        \includegraphics[width=\textwidth]{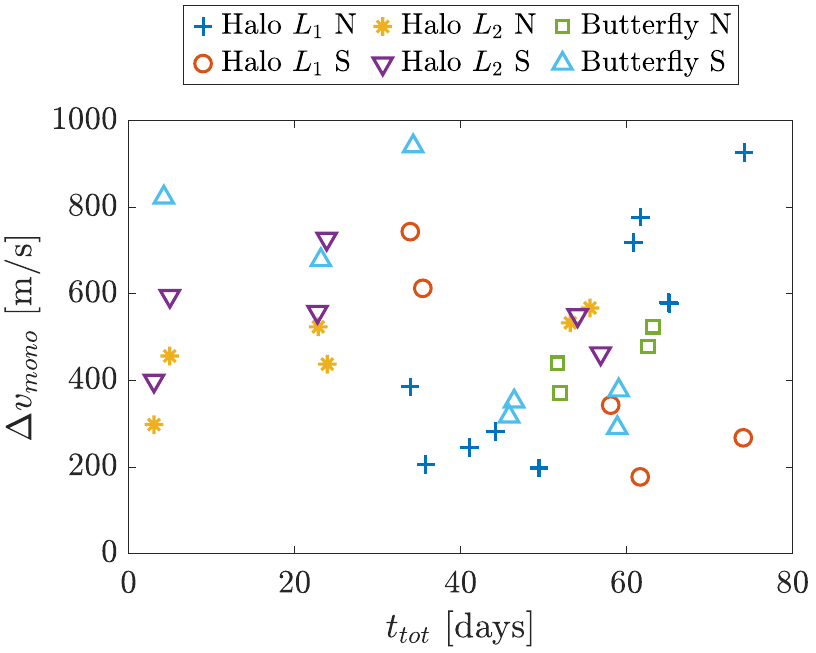}
        \subcaption{Mono-impulsive costs $\Delta v_{\text{mono}}$}
        \label{fig: sample16 Gamma90 dvmono all fams}
    \end{subfigure}
    \hfill
    \begin{subfigure}[t!]{0.48\textwidth}
        \includegraphics[width=\textwidth]{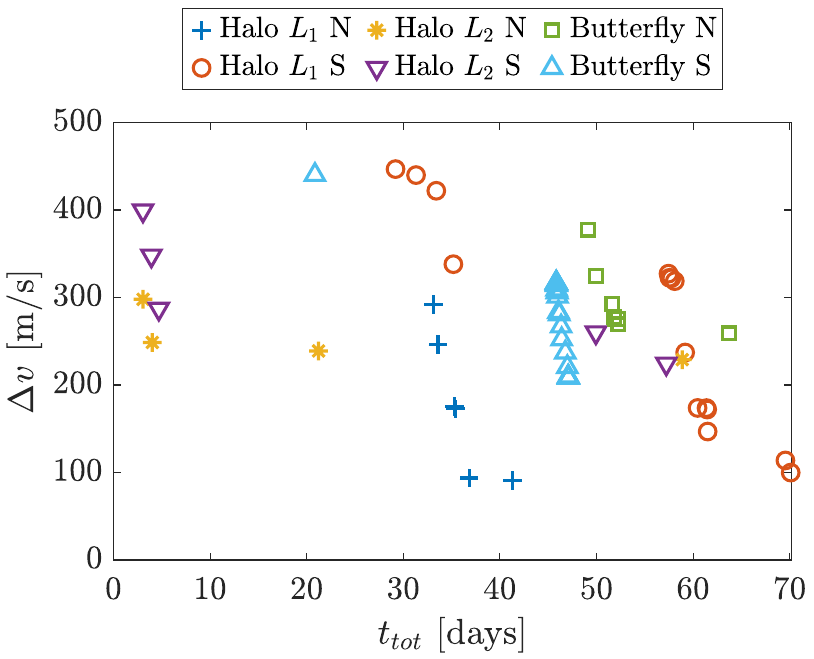}
        \subcaption{Optimal bi-impulsive costs $\Delta v^*$ in the Pareto front}
        \label{fig: sample16 Gamma90 pareto all fams}
    \end{subfigure}
    \caption{Mono- and bi-impulsive cost for transfers from sample \gls{bc}~$\#10/50$ to all the available families.}
    \label{fig: sample16 Gamma90 all fams}
\end{figure}

We now focus on the 10th \gls{bc} from the subset introduced in \cref{sec: subset spatial Gamma90}, referred to as \gls{bc}~$\#10/50$. 
The resulting transfers from this initial condition to all available \gls{po} families are summarized in \cref{fig: sample16 Gamma90 all fams}.
The mono-impulsive cost $\Delta v_{\text{mono}}$ shown in \cref{fig: sample16 Gamma90 dvmono all fams} highlights how the proximity of different families evolves over time $t_{tot}$. This trend becomes even clearer when considering the optimal bi-impulsive costs in \cref{fig: sample16 Gamma90 pareto all fams}. We can infer that, at first, the dynamics of this specific \gls{bc} is partially influenced by the northern halo $L2$ family, as cheap solutions are found for transfers into this family. However, the higher cost compared to subsequent solutions, along with the gap observed for $t_{tot} = 5 \div 20$, suggests that a different family may be dominant at this stage of the capture. This family is not included in the abacus of~\cite{caleb2023}, but may correspond to the one introduced by Aydin \textit{et al.}~\cite{aydin2025POfamilies} as the ``bridge between planar and vertical Lyapunov orbits'', which bifurcates from the Lyapunov family (denoted there as $a$) at point $a^{(1,2)}$.
Around $t_{tot} = 40$ days, the northern halo $L1$ family begins to exhibit a low-cost insertion window. Shortly afterward, the southern butterfly family becomes the most favorable target, although other families quickly start to overlap after that. Toward the end of the capture, the \gls{bc} trajectory closely approaches the dynamics of the southern halo $L1$ family, offering additional low-cost insertion opportunities.
The likely sequence of dominant families influencing this sample \gls{bc}~$\#10/50$ is: northern halo $L2$, the Lyapunov subfamily bifurcating from $a^{(1,2)}$, northern halo $L1$, southern butterfly, and southern halo $L1$. 

An interesting feature observed in \cref{fig: sample16 Gamma90 all fams} is that insertion opportunities into both symmetric subfamilies of the same family (e.g., northern and southern) often emerge nearly simultaneously. Finally, it is noteworthy that each \gls{bc} is associated with at least one accessible \gls{po} family, thereby reinforcing the foundational assumption guiding the initial guesses in the spatial optimization procedure. 

\begin{figure}[tb]
\centering
\includegraphics[width=0.63\textwidth]{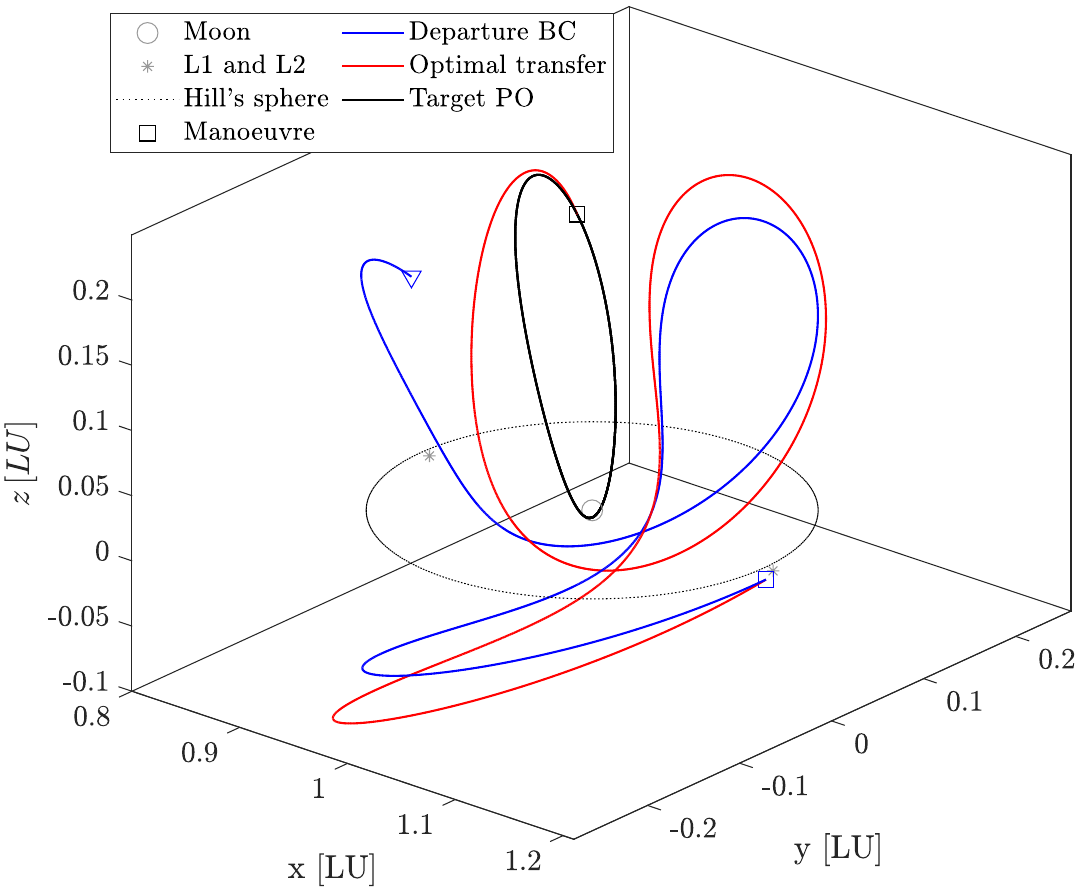}
\caption{Best solution of \cref{fig: sample16 Gamma90 pareto all fams}: transfer from sample \gls{bc}~$\#10/50$ to \gls{nrho} with $\Delta v^* \approx 91$ m/s.}
\label{fig: sample16 to NRHO}
\end{figure}

The best bi-impulsive solution contained in \cref{fig: sample16 Gamma90 pareto all fams} is shown in \cref{fig: sample16 to NRHO}. The cost to insert into the northern halo $L1$ family is approximately $\Delta v^*=91$ m/s. In this specific case, an insertion into \gls{nrho} is achieved, showcasing the potential of the proposed approach for mission design scenarios.


\section{Transfers refinement using convex optimization} \label{sec: refinement convex}

A refinement of the bi-impulsive solutions belonging to the Pareto front for sample \gls{bc}$\#10/100$ targeting the northern halo family around $L1$ is proposed. The refinement is performed using a \gls{scp}~\cite{malyutaConvexOptimizationTrajectory2022} framework, a direct method capable of quickly and efficiently obtaining fixed-time, multi-impulsive trajectories. This implementation utilizes a methodology similar to that presented by Yarndley \textit{et al.}~\cite{Yarndley2025, HoltISSFD}.

Firstly, an appropriate convex linearization for the dynamical system with impulsive maneuvers is obtained. We adapt the dynamics from \cref{eq: equations of motion} by introducing $g(\mathbf{r})=[\ddot{x}, \ddot{y}, \ddot{z}]^T$ and adding an impulsive maneuver $\Delta \mathbf{v}$ at $\tau=\tau_m$:
\begin{align}
    \dot{\mathbf{x}} = f(\mathbf{x}, \Delta \mathbf{v}, \tau_m) = \left\{\begin{array}{l}
            \dot{\mathbf{r}} = \mathbf{v} \\
            \dot{\mathbf{v}} = g(\mathbf{r}) + \delta(\tau - \tau_m) \Delta \mathbf{v} \\
        \end{array}\right.
        \label{eq: equations of motion convex}
\end{align}
where $\delta$ is the Dirac delta function.
As in a direct method, the trajectory is split into $M = 200$ fixed-time segments which are defined by $M{+}1 = 201$ bounding nodes indexed as $m = 0, 1, ..., M$. Each node is associated with a possible impulsive maneuver $\Delta \mathbf{v}_m$. Together, these segments form the multi-impulsive trajectory.

The bi-impulsive transfers from previous sections, being both feasible and near-optimal, serve as effective reference trajectories, with boundary conditions given by the initial and final states:
\begin{equation}
 \mathbf{x}_0 = \mathbf{x}_0 \quad (\text{BC}) \, ,  \qquad \qquad \mathbf{x}_f = \mathbf{x}_M + \left[\mathbf{0}, \Delta \mathbf{v}_M \right]^T \quad (\text{PO}) \, .
  \label{equation_initial_final_states}
\end{equation}
Using the proposed discretization, the linearized dynamic constraints are constructed around the reference trajectories. Specifically, given the reference state and control sequence $(\mathbf{\bar{x}}_m, \Delta \mathbf{\bar{v}}_m)$, a discrete linearized form of the spacecraft dynamics is obtained and enforced as a convex constraint within the \gls{scp} framework:
\begin{equation}
  \forall m \in [0,M-1] : \mathbf{x}_{m+1} = \mathbf{A}_m\mathbf{x}_m + \mathbf{B}_m\Delta \mathbf{v}_m + \mathbf{c}_m \, , 
  \label{equation_dynamics_linear}
\end{equation}
where the matrix $\mathbf{A}_m$ is the \gls{stm}. $\mathbf{A}_m$ and $\mathbf{B}_m$ each represent the changes in the final state $\mathbf{x}_{m+1}$ with respect to the initial state $\mathbf{x}_{m}$ of the same segment and impulsive control $\Delta \mathbf{v}_m$, respectively. Finally, $\mathbf{c}_m$ is the residual vector. As the impulse is applied at the segment start, $\mathbf{B}_m$ is identical to the lower half of $\mathbf{A}_m$. These are calculated by the equations:
\begin{align}
    \mathbf{A}_m &= \left. \left[\frac{\partial}{\partial \mathbf{x}} \int_{\tau_m}^{\tau_{m+1}}\dot{\mathbf{x}}\,\text{d}\tau \right]\right|_{(\mathbf{\bar{x}}_m, \Delta \mathbf{\bar{v}}_m)}\\
    \mathbf{B}_m &= \left. \left[\frac{\partial}{\partial \Delta \mathbf{v}} \int_{\tau_m}^{\tau_{m+1}} \dot{\mathbf{x}}\,\text{d}\tau \right]\right|_{(\mathbf{\bar{x}}_m, \Delta \mathbf{\bar{v}}_m)} \\ 
    \mathbf{c}_m &= \mathbf{\bar{x}}_m - \mathbf{A}_m \mathbf{\bar{x}_m} - \mathbf{B}_m \Delta \mathbf{v}_m.
\end{align}
Rather than using an analytic formulation, the partial derivatives are computed with \gls{ad}, which is directly applied to the initial conditions of a numerical integration solver. The \texttt{Tsit5} numerical integrator is used from the \texttt{DifferentialEquations.jl}~\cite{rackauckasDifferentialEquationsJlPerformant2017} library with absolute tolerance $10^{-10}$ and relative tolerance $10^{-10}$. The \gls{ad} is calculated in forward mode through the use of \texttt{ForwardDiff.jl}~\cite{revelsForwardModeAutomaticDifferentiation2016}. 

To maintain linearization accuracy in the presence of strong cislunar nonlinearities, hard trust region constraints are enforced on the dynamics. They are selected to have a constant size throughout the \gls{scp} algorithm, where
\begin{align} \label{equation_state_trust_regions}
    \forall {m}: -\epsilon_1 \leq \mathbf{x}_m -\bar{\mathbf{x}}_m \leq \epsilon_1 \, .
\end{align}
A range of values for the initial size of the trust regions was tested, and it was found that an $\epsilon_1$ value of approximately $10^{-2}$ tends to provide a good trade-off between convergence and accuracy.

To represent the Euclidean norm of the control inputs within a convex framework, each impulse $\Delta \mathbf{v}_m$ is associated with a scalar auxiliary variable $\Delta v_m$, constrained through a lossless relaxation via a second-order cone (SOC) constraint:
\begin{align} \label{equation_soc_dv}
\Delta v_m \geq \| \Delta \mathbf{v}_m \| \quad (\text{SOC}).
\end{align}
Because we minimize the total $\Delta v$, this constraint is binding at optimality.

The objective of the \gls{scp} is to minimize the total cost, leading to a (convex) optimization problem formulation:
\begin{mini}
    {}{J = \sum_{m=0}^{M} \Delta v_m}{}{}
    \addConstraint{\eqref{equation_dynamics_linear}}{}{\quad\text{(linearized dynamics)}}
    \addConstraint{\eqref{equation_initial_final_states}}{}{\quad\text{(initial and final states)}}
    \addConstraint{\eqref{equation_state_trust_regions}}{}{\quad\text{(state hard trust regions)}}
    \addConstraint{\eqref{equation_soc_dv}}{}{\quad\text{(control magnitude)}}
    \label{equation_full_scp_problem}.
\end{mini}
The \gls{scp} procedure iteratively solves \eqref{equation_full_scp_problem} using a convex solver, updating the linearized dynamics \eqref{equation_dynamics_linear} at each iteration with the latest optimal solution. Convergence is assessed based on the agreement between the linearized dynamics and the true propagated trajectory, which was typically achieved within 30 iterations. The implementation uses \texttt{JuMP.jl}~\cite{lubinJuMPRecentImprovements2023} for problem modeling and MOSEK~\cite{mosek2025} as the convex solver.

While the use of \gls{scp} enables rapid post-processing, refining each bi-impulsive trajectory in under one second on standard hardware, many of the computed $\Delta v$ values are not exactly zero (though effectively negligible, around $10^{-8}$). This is a common problem with direct solvers. In order to address this, a final re-optimization step is performed in which near-zero impulses are fixed to zero. This preserves capture accuracy while having minimal impact on the total $\Delta v$, and the process typically converges within just a few iterations.

\begin{figure}[bt]
    \begin{subfigure}[t!]{0.47\textwidth}
        \centering
        \includegraphics[width=\textwidth]{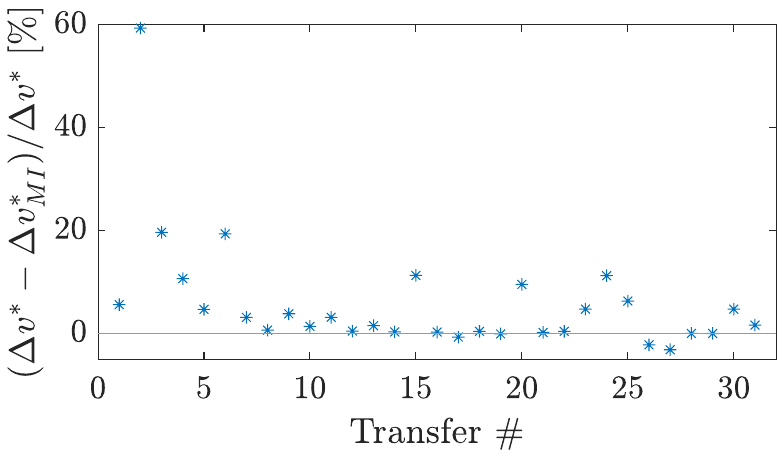}
        \subcaption{Relative (percentage) improvement in $\Delta v^*$}
        \label{fig: comparison convex vs bi-imp dv}
    \end{subfigure}
    \hfill
    \begin{subfigure}[t!]{0.45\textwidth}
        \includegraphics[width=\textwidth]{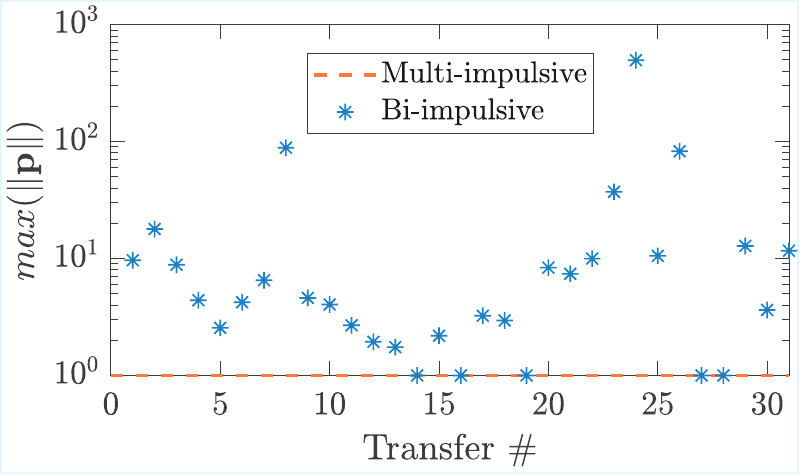}
        \subcaption{Maximum $\|\mathbf{p}\|$ along the transfer}
        \label{fig: comparison convex vs bi-imp PV}
    \end{subfigure}
    \caption{Comparison of bi-impulsive and refined multi-impulsive solutions.}
    \label{fig: comparison convex vs bi-imp}
\end{figure}

\Cref{fig: comparison convex vs bi-imp dv} presents a comparison between the bi-impulsive solutions $\Delta v^*$ and the corresponding refined multi-impulsive solutions $\Delta v^*_{MI}$ obtained via \gls{scp}. The results are expressed as the relative (percentage) improvement of the refined solution: $(\Delta v^* - \Delta v^*_{MI})/\Delta v^*$.
Among all trajectories, only solution~$\#2$ undergoes an important change, with its total cost more than halved after refinement. This behavior arises from the (previously mentioned) occasional slow progress or numerical difficulties of the bi-impulsive optimizer under the highly nonlinear dynamic constraints of the CR3BP, which can lead to premature termination of the optimization process and prevent full exploration of the variable space. 
Approximately one-quarter of the solutions exhibit a substantial improvement in the range of $10\%$–$20\%$, while another quarter shows minor improvement. For the remaining half, the refinement yields negligible change or, in some cases, even a slightly higher cost. These minor increases in cost are not attributable to the convex optimization process but instead arise from the polynomial approximations used in the bi-impulsive transfer computations. As noted in the accuracy checks, deviations within approximately $10$ m/s are considered acceptable. All discrepancies shown in \cref{fig: comparison convex vs bi-imp dv} remain well within this tolerance, with the largest observed difference being only $4$ m/s.
\Cref{fig: comparison convex vs bi-imp PV} compares the maximum primer vector magnitude $\|\mathbf{p}\|$~\cite{jezewski1975PrimerVector} for the two transfer types. The multi-impulsive convex optimization method consistently achieves primer vector optimality, i.e., $\max(\|\mathbf{p}\|)=1$.

\begin{figure}[bt]
    \begin{subfigure}[t!]{0.435\textwidth}
        \centering
        \includegraphics[width=\textwidth]{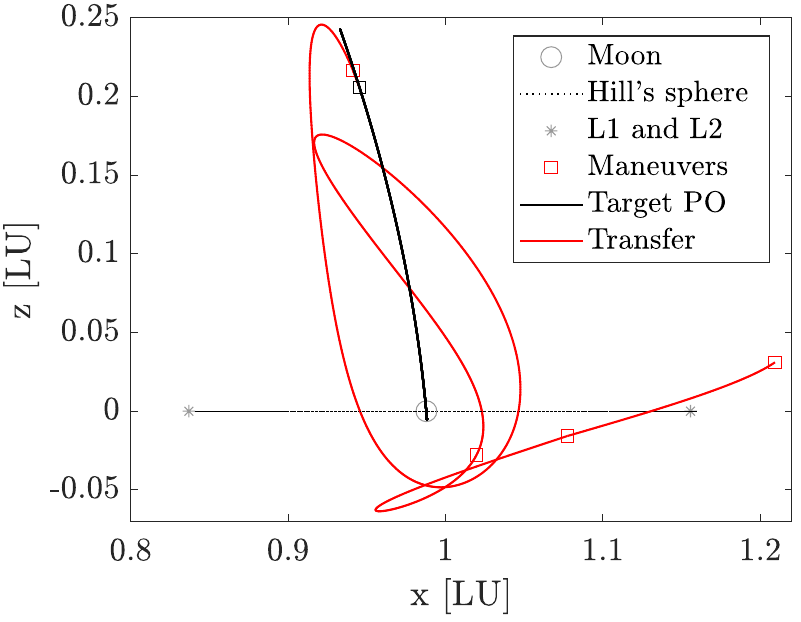}
        \subcaption{$y$-$z$ view}
        \label{fig: convex sample 16 xz}
    \end{subfigure}
    \hfill
    \begin{subfigure}[t!]{0.54\textwidth}
        \includegraphics[width=\textwidth]{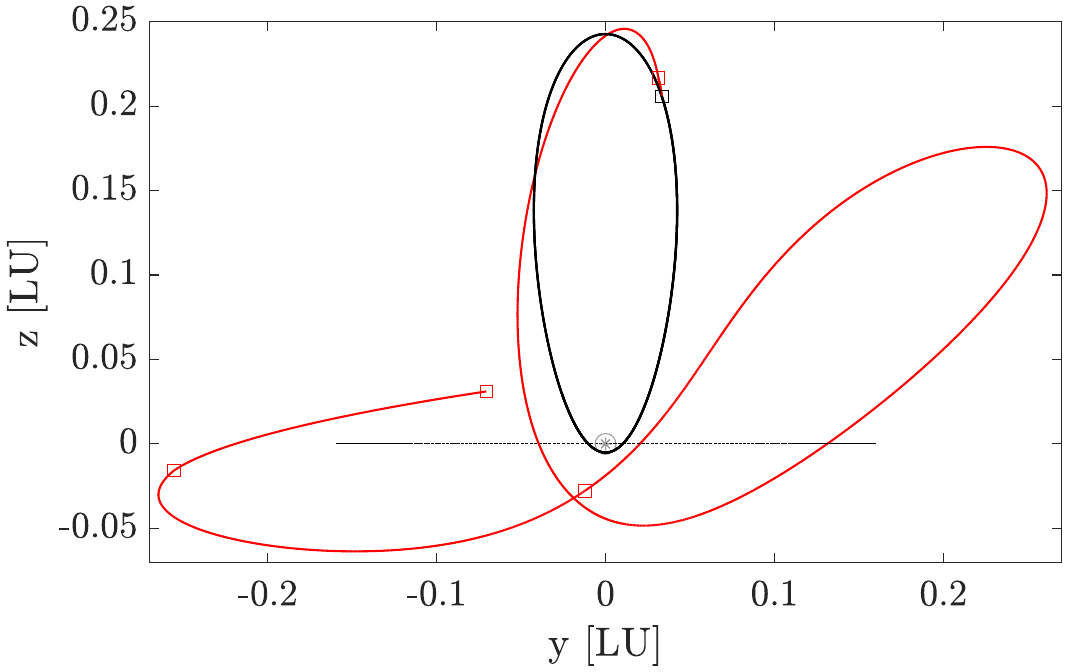}
        \subcaption{$x$-$z$ view}
        \label{fig: convex sample 16 yz}
    \end{subfigure}
    \caption{Fixed-time multi-impulsive convex optimization of sample \gls{bc}~$\#10/100$ to the northern halo $L1$ family.}
    \label{fig: convex sample 16}
\end{figure}

The previously introduced best bi-impulsive transfer from sample \gls{bc}~$\#10/100$ to the northern halo $L1$ family (see \cref{fig: sample16 to NRHO}) achieved a cost of $\Delta v^* \approx 91$ m/s. This corresponds to transfer~$\#6$ in \cref{fig: comparison convex vs bi-imp}, whose multi-impulsive refinement is illustrated in \cref{fig: convex sample 16}.
The refined solution achieves a cost of $\Delta v^*_{MI} \approx 74$ m/s, and remains the lowest-cost transfer even after the convex optimization refinement.

A similar convex optimization framework was previously adopted by Jacini \textit{et al.}~\cite{IppolitaIAC}, who demonstrated the near-optimality of bi-impulsive solutions. Here, we extend their findings by quantifying the degree of suboptimality and confirming that these transfers provide high-quality initial guesses not only for three-impulse~\cite{grossi_optimal3LunarTransfers4BP}, but also for more general multi-impulsive optimizations. While not explored here, the same concept could be applied to low-thrust or higher-fidelity models. Finally, the $L1$ Halo family is selected as a representative and mission-relevant case, particularly suited for applications such as the Lunar Gateway and Lunar Trailblazer missions~\cite{BC-3Djournal}.


\section{Conclusions} \label{sec: conclusions}
\glsresetall

This work presents a high-order optimization framework for computing low-cost transfers from \glspl{bc} to a range of \gls{po} families in the Earth–Moon system. Departure trajectories are drawn from a precomputed database of \glspl{bc}, while the arrivals target \gls{po} families, including distant retrograde orbits (DROs), Lyapunov, halo, and butterfly orbits. By combining differential algebra (DA)-based expansions with polynomial-form constraints on the final state, the method enables accurate and efficient targeting of these \glspl{po}.
Optimization is performed over all relevant parameters, including the Jacobi constant $C_J$ (through the family parameter $p$), enabling flexibility in both the spatial configuration and energy of the final orbit around the Moon. This flexibility is intentional, as it allows the method to probe the dynamical relationship between each \gls{bc} and the surrounding families of \glspl{po}. By identifying which family influences a given capture and when, the approach offers deeper insight into the structure of the phase space, where transfer costs implicitly reflect dynamical proximity. These insights are used to further inform and complement the existing \gls{bc} database and to support the design of low-energy missions that employ a weak stability boundary architecture. In such scenarios, a \gls{bc} can be naturally reached after launch through lunar flybys and solar perturbation, serving as the staging phase preceding the final low-cost orbital insertion investigated here.

The results demonstrate that the proposed method efficiently identifies and optimizes a broad range of viable transfer options. They also demonstrate that the most efficient transfers often correspond to longer transfer times, highlighting the importance of conducting extensive temporal exploration during the design process. When applied to large sets of \glspl{bc}, the method provides insightful statistical characterizations, revealing trends in the transfer options and their dynamical features.
The methodology also proves effective in the spatial case, particularly for targeting near-rectilinear halo orbits (NRHOs), reinforcing its potential utility in mission design contexts such as Gateway and cislunar logistics.
Refinement through convex optimization validates the high-order guesses, producing multi-impulse trajectories with minimal adjustment and confirming their proximity to local optima. These results demonstrate that the proposed approach not only accelerates the search for viable transfers but also yields high-quality candidates suitable for subsequent multi-impulsive refinements. While low-thrust and higher-fidelity dynamical models are not directly addressed in this work, the same framework could be effectively applied to those scenarios in future studies.

\section*{Funding Sources}
The work of Thomas Caleb was funded by SaCLaB (grant number 2022-CIF-R-1), a research group of ISAE-SUPAERO.

\section*{Acknowledgments}
The authors wish to acknowledge the Centre for eResearch at the University of Auckland for their assistance in facilitating this research. http://www.eresearch.auckland.ac.nz



\begin{spacing}{1.2}  
\bibliography{references}
\end{spacing}

\end{document}